\documentstyle[prd,aps]{revtex}

\begin{document}

\draft
\title{Effective Lagrangian Approach to the Theory of Eta
Photoproduction in the $N^{*}(1535)$ Region}
\author{M. Benmerrouche$^{a,b}$,
Nimai C. Mukhopadhyay$^{a,b}
$ and  J. F. Zhang$^a$\\
$^a$ Physics Department\\ Rensselaer Polytechnic Institute\\
Troy, N. Y. 12180-3590 \\
$^{b}$ Saskatchewan Accelerator Laboratory \\ University of Saskatchewan\\
Saskatoon, SK S7N 0W0}

\maketitle

\begin{abstract}
We investigate eta photoproduction in the $N^{*}(1535)$ resonance region
within the effective Lagrangian approach (ELA), wherein  leading
contributions to the amplitude at the tree level are taken into account.
These include the nucleon Born terms and
the leading $t$-channel vector meson exchanges as the non-resonant
pieces. In addition, we consider five resonance contributions
in the $s$- and $u$- channel; besides the dominant $N^{*}(1535)$, these are:
$N^{*}(1440),N^{*}(1520),N^{*}(1650)$ and $N^{*}(1710)$. The amplitudes for
the $\pi^\circ$ and the $\eta$ photoproduction near threshold have significant
differences, even as they share common contributions, such as those of the
nucleon Born terms.  Among these differences, the contribution to the $\eta$
photoproduction of the $s$-channel excitation of the $N^{*}(1535)$ is the most
significant. We find the off-shell properties of the spin-3/2 resonances
to be important in determining the background contributions.
Fitting our effective amplitude to the available data base
allows us to extract the quantity $\sqrt{\chi \Gamma_\eta} A_{1/2}/\Gamma_T$,
characteristic of the photoexcitation of the $N^{*}(1535)$ resonance
and its decay into the $\eta$-nucleon channel, of  interest to
precise tests of hadron models. At the photon
point, we determine it to be $(2.2\pm 0.2)\times 10^{-1} GeV^{-1}$ from the
old data base, and $(2.2\pm 0.1) \times 10^{-1} GeV^{-1}$ from
a  combination of old data base and new Bates data. We obtain
the helicity amplitude for $N^{*}(1535)\rightarrow \gamma p$ to
be $A_{1/2}=(97\pm 7)\times 10^{-3} GeV^{-1/2}$ from the old data base, and
 $A_{1/2}=(97\pm 6)\times 10^{-3} GeV^{-1/2}$ from the combination
of the old data base and new Bates data, compared with the  results of the
analysis of pion photoproduction yielding $74\pm 11$, in the same units.
The observed differential cross-section is not very sensitive to either the
nature of the eta-nucleon coupling or to the precise value of the coupling
constant; we extract a broad range of values
for the $\eta$NN pseudoscalar coupling constant:
$0.2\leq g_{\eta} \leq 6.2$
from our analysis of all available data. We predict, in our ELA,
the angular distributions for a critical series of experiments at Mainz,
and find them  to be in good agreement with the preliminary Mainz data.
Finally, we discuss
implications for future experimental studies with real photons
at the Continuous Electron Beam Accelerator Facility (CEBAF) and other
emerging medium-energy electron accelerators.
Polarization observables, in particular, invite special scrutiny at high
precision.
\end{abstract}



\section{INTRODUCTION }
In recent years there has been a considerable interest\cite{bnl1},
both theoretical and experimental, in the study of the $\eta$ meson and its
interactions with the nucleon.
Suggestions\cite{dover} have been made for using the
$\eta$ to probe $s\bar{s}$ quark component in the nucleon wave
function. Also there is a rising interest in the measurement of
rare and forbidden decays of $\eta$\cite{herczeg} as a test of physics beyond
the standard model.
In the present work, focus is on photoproduction of the eta meson and the
role of production and decay of the $N^{*}(1535)$ resonance (the so-called
$S_{11}$(1535) in the pion-nucleon phase-shift analysis) of spin $1/2$ and odd
parity. This resonance has a remarkably large $\eta N$ branching ratio, a fact
that needs explanation in the theories of hadron structure based on quantum
chromodynamics ($QCD$). It lies only $48MeV$ above the $\eta N$ threshold,
and is the dominant contributor to the photoproduction amplitude even
at threshold. In contrast to the low energy
$\pi N$ and $K N$ interactions, where values of scattering lengths imply a
repulsive interaction\cite{dumbrajs}, the $\eta N$ scattering length obtained
in
the analysis\cite{bhalerao} of $\pi^- p\rightarrow \eta n$ suggests
an attractive interaction between $\eta$ and $N$. More recently, the
analysis of the $pp\rightarrow pp\eta$ near threshold suggests a
possibility of an attractive $\eta NN$ interaction and may also lead to
the formation of ``bound'' $\eta$-nucleus states\cite{chiang}.
  An ideal tool for the
study of the $N^{*}(1535)$ is through electromagnetic processes,  as we
demonstrate in this work.

The eta meson is a member of the ground state SU$(3)$  meson nonet. Thus, the
study of eta photoproduction, the subject of this paper,
shares many of the fundamental
motivations of the extensive study of pion photoproduction
over the past
twenty years or so, including those done at RPI, both
theoretically\cite{dm,dmw,wm,bm,doyle} and experimentally\cite{shaw}.
These  and other related studies have provided an impressive amount of
information about the dynamical properties of the $\Delta(1232)$ as an
isolated quantum mechanical system, and its behavior inside the complex
nucleus. Above this resonance region, however,  large background
contributions\cite{bm1}, and the
overlapping of higher resonances make studying one specific resonance
by pion production mechanisms very difficult. Just as the dominance of the
$\Delta(1232)$ in the $(\gamma,\pi)$ processes have allowed the extraction of
quantitative information on its electromagnetic transition amplitudes,
we hope to extract similar information on $N^{*}(1535)$
via the $(\gamma,\eta)$ process\cite{bm1}. This is an important focus of
this paper.
These photocoupling amplitudes provide useful tests for realistic hadron models
inspired by QCD.

Most of the older data on photoproduction of the eta meson on  protons
come from the
experiments done in the late sixties and early
seventies\cite{delcourt}-\cite{ukai}.
These have been reviewed by
Genzel, Joos and Pfeil\cite{genzel}, and Baldini\cite{baldini}.
The existing older  experimental data are
neither very consistent nor complete in kinematic coverage.
There are large ranges of photon energies and scattering angles,
where no data on
differential cross section exist at present.
The available old  data base on differential cross section on
photoproduction\cite{delcourt}-\cite{ukai} contains
137 points, most of which are for center of momentum (c.m.)
energy below $1.6 \;GeV$.
Only one polarization observable,  the recoil
proton polarization, has been measured\cite{heusch}, but is too poor
(seven data points, of which five are at $90^\circ$) to
be of much theoretical value. To this,  some more  data, of
limited quantity ($15$ differential cross-section data points),
have been added by Homma {\em et al.}\cite{homma} covering
the photon lab energy region from $810$ to $1010MeV$. This data set
has large energy uncertainties of the order $\pm 20MeV$ and the
angular resolution is $\approx \pm 10^\circ$. More recently, members
of the Pittsburgh-Boston-LANL collaboration  at Bates\cite{daehnick,dytman}
have
been able to measure the angular distribution for the ($\gamma,\eta$)
reaction at photon lab energies of $ 729$ and $753 MeV$ at six
angles each. To summarize, the data base for the eta photoproduction have not
reached the relatively high level of accuracy  known for
pion photoproduction in the
$\Delta(1232)$ region. An equivalent multipole set,
crucial in constraining theoretical
models, {\em does not exist}.

With the advent of high-duty cycle electron accelerators, such as
the recently upgraded machines at the Bates (Mass.), Bonn, Mainz, NIKHEF,
and particularly, CEBAF, just coming on-line, systematic and
precise studies of the $\eta$ photoproduction over a wide
range of energies, angles, and momentum transfers should become routine.

First round of experiments at the Mainz Microtron has been
completed, and 45 new data points, as yet preliminary,  have been added
\cite{krusche} to the data base, covering the photon lab energies from
$722\,MeV$ to
$783\,MeV$. This data set has the potential to make the
existing data base almost irrelevant in future, due to better
energy and angular resolutions, and statistics.

Existing theoretical analyses
for the $\gamma p \rightarrow \eta p$ reaction are either
based on a Breit-Wigner type parameterization\cite{homma,hicks,tabakin}
or coupled channel isobar model\cite{bennhold,tiator}. However, it is not clear
how to interpret the couplings extracted in the latter work, as they yield
``bare" couplings that cannot be related to the observable or physical
couplings.
Most of these models have not only suffered from the crudeness of the data, but
also
from the lack of enough theoretical constraints in restricting the number of
parameters fitted, 24 or more.

Recently, there was considerable excitement over the  experiments
at Saclay\cite{saclay} and Mainz\cite{mainz} on the pi-zero photoproduction
near threshold. The results first seemed to indicate a dramatic discrepancy
between the
experimentally determined threshold amplitude\cite{note0}
$E_{0+}$ and the theoretical prediction\cite{debaenst}
based on approximate chiral symmetry (implying a useful Low Energy Theorem,
LET). This
triggered a considerable amount of work\cite{kamal,nozawa,naus,bernard}
on possible corrections to the LET.
However, careful reanalyses\cite{dm,bernstein,bergstrom} concluded that there
were no
significant deviations from the LET prediction for the multipole $E_{0+}$.
Given this intense theoretical discussion on the $\pi^\circ$ photoproduction,
the
$\eta$ photoproduction have taken an added interest\cite{bm1}, in view of the
large
chiral symmetry violation in the case of the eta meson.

After the  Letter by two of us\cite{bm1} on the eta photoproduction has
appeared,
Tiator, Bennhold and Kamalov\cite{tiator} have used a tree-level scheme
motivated by
our work. But there are important differences between our approach and theirs.
First, these authors ignore the complexity of the spin-3/2 particles\cite{bdm},
u-channel resonance exchanges\cite{bm1} etc. Thus, their treatment of
background is
quite different from ours. Second, they  have focused on the extraction of the
eta-nucleon coupling constant.
However, we do not agree
with their primary conclusion that the eta photoproduction data determines the
eta-nucleon coupling quite accurately. This difference between us is not
surprising due to the point (1).
As we show below, the eta-nucleon coupling constant is not
well determined from the data base that we have. We, instead, address here
the  physics of the $N^{*}$(1535).  Finally, due to the multiple scattering
effects, Tiator {\em et al.}\  get bare couplings of the meson-baryon or
photon-baryon vertices that cannot be directly compared to hadron model
predictions, or to those we extract in our paper.

The main objective of this paper is to study photoproduction
of the eta meson on protons from threshold through the $N^{*}(1535)$
resonance region in the effective Lagrangian formalism, with a view to
extract the product of
the $N^{*}(1535)\leftrightarrow p$ electromagnetic transition
amplitude and the amplitude for the decay of the $N^{*}(1535)$
via the eta-nucleon channel, from the existing data.
The remainder of this paper is organized as follows: Section II is concerned
with the
formalism of photoproduction of the eta meson. Kinematics, invariant
amplitudes and
multipole expansion are reviewed.
Section III introduces the effective Lagrangian formalism and the
amplitudes arising from various particle exchanges are given in the tree
approximation.
This section is devoted to the photoproduction mechanism. Some important
theoretical issues  associated with the treatment of the spin-$3/2$
baryons are also examined in this section. Results, followed by a comparison
between
$\pi^\circ$ and $\eta$ photoproduction near threshold, are given
in the section IV.
Section V summarizes our conclusions and poses some future
research problems in this area. Appendices provide some details
of the theory, helpful for further understanding of the formalism.

In brief,  this work explores the tree-level structure of the eta
photoproduction in
the $N^{*}(1535)$ region.
The overwhelming  dominance of the
$N^{*}(1535)$ resonance, demonstrated below, should make the main results
obtained in this paper, substantially immune from the unitarity corrections,
ignored here. The reason for this optimism is that the Breit-Wigner form
of the $N^{*}(1535)$ is already unitary, and other contributions are too
small to be of great consequence in the unitarity violation.

\section{GENERAL FORMALISM}

In order to fix the notation and the convention\cite{note1}
basic formulae for kinematics, invariant amplitudes, differential cross section
and
other observables are reviewed in this section.

\subsection{Kinematics}

In this work, the following reaction of photoproduction of
the $\eta$ meson on a nucleon is  considered

\begin{equation}
\gamma (k) + N(p_i) \rightarrow \eta(q) + N(p_f), \label{eq:1}
\end{equation}
where for each particle we have indicated the four-momentum in
parentheses.

Use is made of the usual invariant quantities (Mandelstam variables):
\begin{eqnarray}
s&=& (k+p_i)^2=(q+p_f)^2,\nonumber \\
u&=& (k-p_f)^2=(q-p_i)^2,  \\
t&=& (q-k)^2=(p_i-p_f)^2, \nonumber
\end{eqnarray}
subject to the constraint $s+t+u=2 M^2+\mu^2+k^2$,
where $M$ and $\mu$ denote the masses of the nucleon and the eta meson.
It is  easier
to work in the  c.m. frame of
the final nucleon and the $\eta$ meson where the experimental
observables are calculated.
The relations (2), in the c.m. system, become
\begin{eqnarray}
s&=& W^2 = (E_i + k_0)^2, \\
u&=& M^2 + k^2 - 2\,k_0\,E_f - 2\,|\vec{q}|\,|\vec{k}| x,\\
t&=& \mu^2 + k^2 - 2\,k_0\,\omega + 2\,|\vec{q}|\,|\vec{k}| x,\\
x&=&\cos\theta = \frac{\vec{q}\cdot\vec{k}}{|\vec{q}|\,|\vec{k}|},
\end{eqnarray}
with $\theta$ being the c.m. scattering angle and $W$, the total c.m. energy.
It is straightforward to derive
the following energies and momenta in terms of $W$ and $k^2$:
\begin{eqnarray}
k_0=\frac{W^2+k^2-M^2}{2W},&\;\;&E_i=\frac{W^2-k^2+M^2}{2W}, \\
\omega=\frac{W^2+\mu^2-M^2}{2W},&\;\;&
E_f=\frac{W^2-\mu^2+M^2}{2W}, \\
|\vec{p}_f|=|\vec{q}|=\sqrt{\omega^2-\mu^2}&=&
\frac{\sqrt{[(W-M)^2-\mu^2][(W+M)^2-\mu^2]}}{2W}, \\
|\vec{p}_i|=|\vec{k}|=\sqrt{k_0^2-k^2}&=&
\frac{\sqrt{[(W-M)^2-k^2][(W+M)^2-k^2]}}{2W},
\end{eqnarray}
where
\begin{equation}
k=(k_0,\vec{k}),\;\;\;p_i=(E_i,-\vec{k}),\;\;\;q=(\omega,\vec{q}),
\;\;\;p_f=(E_f,-\vec{q}).
\end{equation}
For photoproduction,
$k^2=0$, and the relation between the energy $E_\gamma$ of
the photon in the lab frame and the c.m. energy is
\begin{equation}
E_\gamma=\frac{W^2-M^2}{2M}.
\end{equation}
The threshold for the photoproduction of the eta meson is at the
photon lab energy of $709.3 MeV$, corresponding to a  $W$ of $1487.1 MeV$.
These contrast with the corresponding values for the neutral pion of
$144.7$ and $1079.1 MeV$.

\subsection{Invariant amplitudes}

The S-matrix elements for the elementary processes (1)  can
be written as\cite{donnachie}
\begin{equation}
S_{fi}=\frac{1}{(2\pi)^2} \delta^4(p_f+q-p_i-k)
\sqrt{\frac{M^2}{4\omega kE_iE_f}} i{\cal M}_{fi}.
\end{equation}
The invariant matrix element $i{\cal M}_{fi}$ can be decomposed as
\begin{equation}
i{\cal M}_{fi}=\bar{U}(p_f) \varepsilon_\mu O^\mu U(p_i),
\end{equation}
where $U(p_i),\bar{U}(p_f)$ are the Dirac spinors for the initial and
final nucleon respectively; $O^\mu$ describes the current operator produced
by the strongly interacting
hadrons, and $\varepsilon_\mu$ is the photon polarization vector.

The spin dependence can be made explicit by decomposing the hadron current
operator $O^\mu$ in terms of eight most general Lorentz covariant
pseudovectors;
\begin{equation}
O^\mu=\sum_{j=1}^{8} B_j(s,t,u,k^2) N_j^\mu.
\end{equation}
Because of the pseudoscalar nature of the $\eta$ meson, the only Dirac
matrices that enter in (15) are $\gamma_5$, $\gamma_5\gamma_\mu$ and
$\gamma_5\gamma_\mu\gamma_\nu$. Taking into account the Dirac equation
for the incoming and outgoing nucleon, both on shell,
\begin{eqnarray}
\gamma\cdot p_i U_i &=& MU_i,\\
\bar{U}_f \gamma\cdot p_f &=& M\bar{U}_f,
\end{eqnarray}
and conservation of the four-momentum ($p_i+k=p_f+q$),
one can form eight
Lorentz pseudovectors $\bar{U}_{f}N^{\mu}_{j}U_{i}$, where $N_{j}^{\mu}$ are
\begin{equation}
\begin{array}{cccc}
N_1^\mu=\gamma_5 P^\mu,  &
N_2^\mu=\gamma_5 q^\mu,  &
N_3^\mu=\gamma_5 k^\mu,  &
N_4^\mu=\gamma_5 \gamma^\mu, \\
N_5^\mu=\gamma_5 \gamma\cdot k P^\mu, &
N_6^\mu=\gamma_5 \gamma\cdot k q^\mu, &
N_7^\mu=\gamma_5 \gamma\cdot k k^\mu, &
N_8^\mu=\gamma_5 \gamma\cdot k \gamma^\mu,
\end{array}
\end{equation}
where  $P^\mu=\frac{1}{2}(p_i+p_f)^\mu$.
Any other pseudovector can be reduced to a linear
combination of the $N_j^\mu$. Current conservation (gauge invariance)
condition,
$k_\mu j^\mu=0$, and the identity $k^{2}=0$
reduce the number of independent amplitudes to four for
photoproduction\cite{dennery},
where $j^{\mu}$ is the electromagnetic hadron current.
These yield the well-known
Chew-Goldberger-Low-Nambu\cite{cgln} (CGLN) amplitudes.
The matrix element $i{\cal M}_{fi}$ is expanded as
\begin{equation}
i{\cal M}_{fi}= \bar{U}_f(p_f) \sum_{j=1}^{4} A_j(s,t,u,k^2) M_j
U_i(p_i),
\end{equation}
where
\begin{equation}
\begin{array}{rcl}
M_1&=&-\frac{\textstyle 1}{\textstyle 2}
\gamma_5 \gamma_\mu \gamma_\nu F^{\mu\nu}, \\[0.5em]
M_2&=&+2 \gamma_5 P_\mu ( q_\nu -\frac{1}{2}k_{\nu})
F^{\mu\nu}, \\[0.5em]
M_3&=&-\gamma_5 \gamma_\mu q_\nu F^{\mu\nu}, \\[0.5em]
M_4&=&-2\gamma_5 \gamma_\mu P_\nu F^{\mu\nu}-2 M M_1,
\end{array}
\end{equation}
with the electromagnetic field tensor
\begin{equation}
F^{\mu\nu}=\varepsilon^\mu k^\nu-\varepsilon^\nu k^\mu.
\end{equation}
For electroproduction of eta, the sum in Eq. (19) is extended to include
$M_5$ and $M_6$, given by
\begin{eqnarray}
M_5 & = & \gamma_{5}k_{\mu}q_{\nu}F^{\mu\nu} \nonumber\\
M_6 & = & -\gamma_{5}k_{\mu}\gamma_{\nu}F^{\mu\nu} \eqnum{20$'$}
\end{eqnarray}

The above particular form of the invariant amplitudes exhibits
simple properties under crossing symmetry,
when the initial and final nucleon are interchanged\cite{donnachie,berends}.
For the processes under consideration, the crossing (i.e exchange
$s\leftrightarrow u$)
properties of the $A's$ can be readily deduced\cite{donnachie,berends}:
\begin{equation}
\begin{array}{rr}
A_j(s,t,u,k^2)=+A_j(u,t,s,k^2)&  (j=1,2,4),\\
A_j(s,t,u,k^2)=-A_j(u,t,s,k^2)&  (j=3, 5, 6),
\end{array}
\end{equation}
where we have included  also the electroproduction case.
The isospin decomposition of the amplitudes is accomplished in the
following way. The photon interaction has an isovector part  and an
isoscalar part, assuming that it has no isotensor part\cite{donnachie1}.
The vector part leads to isovector amplitudes $A_j^V$ and the scalar
part gives isoscalar amplitudes $A_j^S$. Since the $\eta$ meson
is isoscalar,
only isospin $\frac{1}{2}$ final states are allowed. Thus, the isospin
decomposition of the amplitudes has the simple form
\begin{equation}
A_j=A_j^S+A_j^V\tau_3\;\;\;\;\;\;j=1,...,6,
\end{equation}
where the two physical
amplitudes are given by
\begin{equation}
\begin{array}{cc}
\gamma p\rightarrow \eta p: &A_j^p=A_j^S+A_j^V,\\
\gamma n\rightarrow \eta n: &A_j^n=A_j^S-A_j^V.
\end{array}
\end{equation}
\subsection{The CGLN amplitudes}

It is convenient to reexpress the
invariant amplitudes in terms of amplitudes corresponding to a definite
parity and angular momentum state. The matrix elements
appearing in Eq.(19) are first written in a two-component form by
expressing the
$\gamma$-matrices in terms of the Pauli $\sigma$-matrices, and the
Dirac spinors in terms of the two component spinors:
\begin{equation}
{\cal M}_{fi}=\frac{4\pi W}{M} \chi_f^\dagger {\cal F} \chi_i,
\end{equation}
where the $\chi_i,\chi_f$ are the initial and final nucleon
Pauli spinors and the factor $\frac{\textstyle 4\pi W}{\textstyle M}$
has been introduced as a definition of the ${\cal F}$.
It can be written in the familiar form\cite{berends}
\begin{equation}
\begin{array}{rcl}
{\cal F}&=&i\vec{\sigma}\cdot \vec{b} {\cal F}_1 +
\vec{\sigma}\cdot\hat{q} \vec{\sigma}\cdot (\hat{k}\times\vec{b})
{\cal F}_2 + i\vec{\sigma}\cdot\hat{k}\hat{q}\cdot\vec{b} {\cal F}_3  \\
&&+i\vec{\sigma}\cdot\hat{q}\hat{q}\cdot\vec{b} {\cal F}_4,
\end{array}
\end{equation}
with $\vec{b} = \vec{\varepsilon}$ for photoproduction. For electroproduction,
\begin{equation}
b_\mu=\varepsilon_\mu-\frac{\vec{\varepsilon}\cdot\hat{k}}{|\vec{k}|}k_\mu,
\;\;\hat{k}=\frac{\vec{k}}{|\vec{k}|},\;\;\hat{q}=\frac{\vec{q}}{|\vec{q}|},
\end{equation}
and two extra terms are to be added to $\cal F$ in
Eq. (26). This extra piece is
$$-i\vec{\sigma} \cdot\hat{q}b_{0}{\cal F}_{5}-i\vec{\sigma}\cdot\hat{k}b_{0}
{\cal F}_{6}.$$
Note that $\vec{b}\cdot\vec{k}=0$, so that $b_\mu$ has no longitudinal
component.
The relations between
the ${\cal F}'s$ and the $A's$ is derived by reducing Eq.(19) into
two-component spinors. One obtains
\begin{equation}
\begin{array}{rcl}
{\cal F}_1&=&\frac{\textstyle ab}{\textstyle 8\pi W}
\{(W-M)A_1+(W-M)^2 A_4+q\cdot kA_{34}\},  \\[0.5em]
{\cal F}_2&=&\frac{\textstyle qk}{\textstyle 8\pi Wab}
\{-(W+M)A_1+(W+M)^2 A_4+q\cdot kA_{34}\},   \\[0.5em]
{\cal F}_3&=&\frac{\textstyle qkb}{\textstyle 8\pi Wa}
\{(W^2-M^2)A_2+(W+M)A_{34}\},  \\[0.5em]
{\cal F}_4&=&-\frac{\textstyle q^2 a}{\textstyle 8\pi Wb}
\{(W^2-M^2)A_2-(W-M)A_{34}\},
\end{array}
\end{equation}
where
\begin{eqnarray*}
&A_{34}&=A_3-A_4.
\end{eqnarray*}
For electroproduction, ${\cal F}_{1}$, ${\cal F}_{2}$ will have the following
extra terms inside the brace:$-k^2 A_6$, and ${\cal F}_{3}$, ${\cal F}_{4}$
will have $-k^2 A_2 /2 +k^2 A_5$. And there will be extra amplitudes
${\cal F}_5$, ${\cal F}_6$ given by
\begin{eqnarray}
{\cal F}_5 & = & -\frac{qa}{8\pi Wb}\{ (E_i -M)[ A_{1}-(W+M)A_{46}]
+[ q\cdot k-\omega (W-M)]A_{34}-A_{25}\}, \nonumber \\
{\cal F}_6 & = & \frac{kb}{8\pi Wa}\{(E_i +M)[A_1 +(W-M)A_{46}]-
[q\cdot k-\omega (W+M)]A_{34}-A_{25}\},
\eqnum{28$'$}
\end{eqnarray}
where $$A_{25}=[\vec{q}\cdot\vec{k}(\frac{3}{2}k_{0}-2W)+|\vec{k}|^{2}(W-
\frac{3}{2}\omega )]A_2 -(\omega k^2 -k_0 q\cdot k )A_5, $$
$$A_{46}=A_4 -A_6 .$$
Here the notation $k=|\vec{k}|,\;q=|\vec{q}|$ has been adopted and
the definitions $a=\sqrt{E_i+M}$ and $b=\sqrt{E_f+M}$ have been used.
By a tedious but straightforward manipulation\cite{goldberger}, one can relate
${\cal F}_i$ to the multipoles, which are classified according to the nature of
the
photon and the total angular momentum
$J=\ell\pm\frac{1}{2}$ of the final state. The transverse photon states
can be electric, $EL$, with parity $(-)^L$, or magnetic, $ML$,
with parity $(-)^{L+1}$,
where $L$ is the total angular momentum of the photon, $L \geq 1$.
The scalar (or longitudinal) photons are relevant for the electroproduction
of mesons: the corresponding multipoles are $SL$, with parity $(-)^L$.
The $N^{*}(1535)$ resonance with
spin $J=\frac{1}{2}$ and negative parity, corresponding to $\ell =0$
of $\eta N$ final state, can be excited, with real photons, by the $E1$
radiation. The corresponding multipole would be $E_{0+}$ .
For the electroproduction of etas, we would additionally have
the scalar multipole $S_{0+}$. In the Section
III, we shall discuss the non-resonant (Born) and the s-channel N*(1535)
contributions to this multipole, and its difference with that in the case
of the pi-zero photoproduction.

\section{PHOTOPRODUCTION OF THE $\eta$ MESON}

The effective Lagrangian approach\cite{dmw,olsson,peccei} helps
us sort out the tree-level structure (Fig. 1) of the $\eta$
 photoproduction
amplitude by considering the leading exchanges in the s, t and u
channels. The procedure is exactly parallel to the pion
photoproduction. The leading $s$- and $u$-channel nucleon
Born terms, along with vector meson in the $t$ channel, are added to the
contribution of nearby resonances in the $s$- and $u$-channel.

\subsection{Nucleon Born terms}

In pion photoproduction, the $\pi NN$ strong interaction vertex is  treated
phenomenologically using an effective Lagrangian\cite{peccei}. The two
standard ways of introducing the pion-nucleon interaction are via the
pseudoscalar ($PS$) or pseudovector ($PV$) couplings. For elementary
fields, without anomalous magnetic interactions,
the  $PS$ and $PV$ Lagrangians are equivalent, in the
lowest order in strong coupling constant. Anomalous magnetic moments of the
nucleon are the
reason for the breaking for this equivalence, as discussed below.
It is well known that the amplitudes derived from the $PV$ coupling
are in accord with the low energy theorem ($LET$) based on gauge
invariance and (approximate) chiral
symmetry\cite{debaenst}. However, because of the relatively large $\eta$ mass
and big
breaking of the chiral $SU(3)\times SU(3)$
symmetry\cite{gellmann,manohar,meissner,maltman},
compared to chiral $SU(2)\times SU(2)$ one\cite{gasser,bernard1}, there is no
compelling reason to choose the $PV$ form of the $\eta NN$ coupling. The
Feynman diagram for PV amplitudes are shown in Fig. 1(a)-(c), where (a)
and (b) represent the usual $PS$ amplitudes and (c) is the seagull diagram
(proportional to the nucleon anomalous magnetic moment)
for the equivalence breaking term, not to be confused with
the traditional $PV$ seagull term. Therefore the difference between the $PS$
and the $PV$ coupling at the tree level arises from the fact that the
nucleon is a composite particle.
Since $\eta$ is a neutral hadron, the $t$-channel $\eta$
exchange and the $PV$ seagull contribution are absent.
The effective $\eta NN$ interaction Lagrangian can be written as\cite{gross}
\begin{equation}
{\cal L}_{\eta NN}=g_\eta [-i\zeta\bar{N}\gamma_5
N\eta+(1-\zeta)\frac{1}{2M}\bar{N}\gamma_\mu \gamma_5 N\partial^\mu \eta],
\end{equation}
$M$ being the nucleon mass, the two limiting cases being $\zeta=0\;(PV)$ and
$1\;(PS)$, the coupling $g_\eta$ for the $\eta NN$
vertex is not very well-known\cite{dumbrajs} and  several methods have been
used to determine its
magnitude. In the $SU(3)$ flavor model,
the value of $g_{\eta_8}$, where
$\eta_8$ refers to the pure $SU(3)$ octet eta meson, is related to the
well-known $\pi NN$ coupling constant $g_\pi$ through the
relation\cite{dumbrajs}
$$g_{\eta_8}=\frac{1}{\sqrt{3}} (3-4\alpha_p)g_\pi,$$
where $\alpha_p=D/(D+F)$ with $D$ and $F$ being the two type of $SU(3)$
octet meson-baryon couplings. It should be remembered
that $\eta_8$ does not accurately represent the physical $\eta$,
which, though predominantly in an $SU(3)$ octet, contains an  admixture of the
$SU(3)$ singlet configuration. Since the admixture is small\cite{genz,imz},
we shall ignore it to estimate here the $g_\eta$, to be taken as $g_{\eta_8}$.
Values for $\alpha_p$ range between $0.59$ and $0.66$\cite{peng,nagels},
which gives
$g_\eta^2/g_\pi^2$ between $0.043$ and $0.14$.
Other methods of its determination, such as the ratio of the
backward angle cross sections of the reactions
$\pi^- p\rightarrow \eta n$ and $\pi^- p\rightarrow \pi^\circ n$ (or
$K^- p\rightarrow \eta \Lambda$ and $K^- p\rightarrow \pi^\circ
\Lambda$), have given a range of $0.18$ to $0.45$ for
$g_\eta^2/g_\pi^2$\cite{peng}.
The $\eta$ meson coupling has also been determined from fits to low energy
nucleon-nucleon\cite{machleidt} scattering in the one boson exchange models,
yelding  $g_\eta^2/g_\pi^2\simeq 0.35$.
Finally, the $SU(6)_W$\cite{gupta} symmetry gives $g_\eta=\sqrt{3}g_\pi/5$,
which implies
$g_\eta^2/g_\pi^2=3/25$, comparable to the $SU(3)$ value given above.
Taking $g_\pi^2/4\pi=13.4$, a conservative range of values for
$g_\eta$ is:
\begin{equation}
0.6\leq g_\eta^2/4\pi \leq 6.4.
\end{equation}
Thus, the coupling for the $\eta NN$ vertex is  uncertain, but significantly
smaller than
the $\pi NN$ coupling. In our work, the $\eta NN$ coupling will be allowed to
vary within a range of values bounded by zero and $6.4$, unless stated
otherwise.

The $\gamma NN$ interaction Lagrangian is well-determined within the
framework of the  quantum electrodynamics (QED):
\begin{equation}
{\cal L}_{\gamma NN}=-e \overline N \gamma_{\mu} {{(1+\tau_3)}\over 2} N
A^{\mu}+{e\over{4M}} \overline N (k^s+k^v \tau_3)
\sigma_{\mu \nu}N F^{\mu \nu}.
\end{equation}
$\alpha = e^{2}/4\pi  \simeq 1/137$,
$N,\eta,A^\mu$ are the nucleon, eta and photon fields, $k^s$ and $k^v$
are the isoscalar and isovector nucleon anomalous
magnetic moments, $k^s={1 \over 2}(k_p+k_n), \;\;\ k^v={1 \over
2}(k_p-k_n)$,
with  $k_p=1.79$ nuclear magnetons (nm) and $k_n=-1.91$ nm, $F^{\mu
\nu}=\partial^{\nu}A^{\mu}-\partial^{\mu}A^{\nu}$.
To a first order in $e$ and $g_\eta$, the interaction Lagrangians (29) and (31)
yield,
for the $PS$ coupling ($\zeta=1$), the nucleon exchange amplitude:
\begin{eqnarray}
i{\cal M}_{fi}^{PS}&=&eg_\eta \bar{U}_f
\left [\gamma_5
\frac{\gamma\cdot (p_i+k)+M}{s-M^2}
\{\frac{(1+\tau_3)}{2}\gamma\cdot\varepsilon +
\frac{(k^s+k^v \tau_3)}{2M} \gamma\cdot k\gamma\cdot\varepsilon\} \right.
\nonumber \\
&&\left. +\{\frac{(1+\tau_3)}{2} \gamma\cdot\varepsilon +
\frac{(k^s+k^v \tau_3)}{2M} \gamma\cdot k\gamma\cdot\varepsilon\}
\frac{\gamma\cdot (p_f-k)+M}{u-M^2} \gamma_5 \right ] U_i,
\nonumber \\
&&
\end{eqnarray}
whereas for the $PV$ case ($\zeta=0$),
$\gamma_5$ should be replaced by $\gamma_5 \gamma\cdot q$ in the above
expression. The first term on the right hand side of Eq. (32) is due to the $s$
channel diagram, Fig. 1(a), while the second one corresponds to the
$u$-channel diagram, Fig. 1(b). In the case of a neutron
target and the reaction ($\gamma n\rightarrow \eta n$), the charge terms
proportional to
$(1+\tau_3)$ vanish.

An alternative way to introduce $PV$ coupling is to transform
the original $PS$ interaction Lagrangian by redefining
the nucleon field through the chiral rotation operator
$U=\exp (i\frac{g_\eta}{2M}\gamma_5 \eta)$\cite{friar}.
To first order in $g_\eta$ in the tree approximation,
modifying the nucleon field in the total $PS$ Lagrangian (including the free
one) by
\begin{equation}
N\rightarrow N+i\frac{g_\eta}{2M}\gamma_5 \eta N,
\end{equation}
leads to
\begin{equation}
{\cal L}^{PV}={\cal L}^{PS}+i\frac{ef_\eta}{2M\mu}
\bar{N}(k^s+k^v \tau_3)\sigma_{\mu\nu}\gamma_5 NF^{\mu\nu} \eta,
\end{equation}
where $f_\eta/\mu=g_\eta/(2M)$. The last term on the right hand side of
Eq.(34) is the equivalence breaking term (also known as the
"catastrophic" term\cite{friar}) and is depicted in Fig. 1(c). Its
corresponding amplitude is given by:
\begin{equation}
i{\cal M}_{fi}^{EB}=\frac{ef_\eta}{M\mu}
\bar{U}_f (k^s+k^v \tau_3)\gamma\cdot\varepsilon\gamma\cdot k \gamma_5 U_i.
\end{equation}
It is straightforward now to determine the $A's$ appearing in
Eq.(19),
\begin{eqnarray}
A_1^{PS}&=&e_N e g_\eta [\frac{1}{s-M^2}+\frac{1}{u-M^2}], \\
A_2^{PS}&=&2e_N e g_\eta \frac{1}{(s-M^2)(u-M^2)}, \\
A_3^{PS}&=&-e g_\eta \frac{k_N}{2M}[\frac{1}{s-M^2}-\frac{1}{u-M^2}], \\
A_4^{PS}&=&-e g_\eta \frac{k_N}{2M}[\frac{1}{s-M^2}+\frac{1}{u-M^2}],
\end{eqnarray}
and according to Eq.(34), the $PV$ $A's$ are given by
\begin{eqnarray}
A_1^{PV}&=&A_1^{PS}+e g_\eta \frac{k_N}{2M^2},\\
A_j^{PV}&=&A_j^{PS},\;\;\;\;\;\; j=2,3,4,
\end{eqnarray}
where $e_N=e_p=+1,k_N=k_p$ for protons and
$e_N=e_n=0,k_N=k_n$ for neutrons. The CGLN amplitudes can now be
determined using Eq.(28), which lead to the multipoles (given in
Eq.(B9)).

  There are a few remarks  concerning this
part of the theory:
(1) The nucleon Born terms will project in all
the multipoles, including the dominant multipole $E_{0+}$.
(2) The equivalence breaking
term contributes only to $E_{0+}$ and $M_{1-}$, and therefore the
difference between the $PS$ and $PV$ couplings would only appear into these
multipoles.
(3) In our phenomenological fits, the parameter $g_\eta$ is allowed to
vary, while $\zeta$ is fixed to either $1$ or $0$, corresponding to
$PS$ or $PV$ cases.
(4) In principle, one should attach form factors at
various vertices in Fig. 1(a)-1(b), since the intermediate nucleon is
off-shell. This, in general, may not preserve gauge
invariance\cite{note2}
and a special
procedure needs to be implemented to maintain gauge invariance.
Given the poor quality and scarcity of the data,
and the general agreement with data, as discussed
later, form factors are not introduced at the nucleon vertices.

\subsection{Vector meson exchanges}

Though the inclusion of the t-channel exchanges, together with the
contributions from other channels may, in general, clash  with the
ideas of duality\cite{schwela},
their r\^{o}le is unmistakable
in the dispersion theory\cite{berends}. The r\^{o}le of the t-channel
vector meson exchange in neutral pion photoproduction\cite{dmw,olsson} is
clearly indicated.
Thus, they would be included here to see if these contributions influence our
ability
to  extract the nucleon to $N^{*}(1535)$ electromagnetic amplitude.
The strong and electromagnetic vertices involving the vector meson
[Fig. 1(d)] are described by the Lagrangians\cite{olsson}:
\begin{eqnarray}
{\cal L}_{VNN}&=&-g_v {\overline N} \gamma_{\mu}NV^{\mu}+
       {g_t\over{4M}}{\overline N} \sigma_{\mu \nu}NV^{\mu \nu}, \\
{\cal L}_{V\eta\gamma}&=&\frac{e\lambda_V}{4\mu} \epsilon_{\mu \nu \lambda
\sigma}F^{\mu \nu}V^{\lambda \sigma} \eta,
\end{eqnarray}
with the vector meson field tensor
$V_{\mu \nu}=\partial_{\nu}V_{\mu}-\partial_{\mu}V_{\nu}$,
$\mu$ being the eta mass.
In this energy region, it is enough to consider $\rho$ and $\omega$
mesons. The r\^{o}le of $\phi$ meson is found to be unimportant (less
than two percent of the $\rho+\omega$ contribution at threshold), not
surprising in the light of the Okubo-Zweig-Iizuka suppression.
Exchanges from heavier mesons are expected to be negligible, because of
their smaller $\gamma \eta$ decay widths and larger masses.
Since the $\eta$ meson is  a neutral particle, in the lowest order,
the t channel $\eta$ diagonal exchange contribution is zero.
Contribution arising from anomalies and connected
with the process $\eta \rightarrow 2 \gamma$,
is small and is neglected.  Some
properties of the vector mesons, pertinent to this work, are summarized
in Table \ref{table1}.
The vector and tensor couplings (see Table \ref{table1}) of the vector
meson-nucleon
vertex are taken from analyses of strong interaction processes such as $\pi N$
and $N N$
scattering using dispersion relations\cite{hoehler,grein}. Using the Lagrangian
${\cal L}_{V\eta\gamma}$, it is straightforward to calculate the decay
width $V\rightarrow \gamma \eta$, which is related to
the radiative coupling $\lambda_V$ by
\begin{equation}
\Gamma_{V\rightarrow \gamma \eta}=
\frac{\alpha (M_V^2-\mu^2)^3}{24 \mu^2 M_V^3} \lambda^{2}_V.
\end{equation}
 From the study of radiative decays\cite{dolinsky}, extracted parameters,
given in Table \ref{table1}, yield
\begin{equation}
\lambda_\rho=1.06 \pm  0.15,\;\;\;\;\;\;\lambda_\omega=0.31 \pm 0.06,
\end{equation}
in good agreement with quark model calculation\cite{godfrey} which predicts
$\lambda_\rho\simeq 3\lambda_\omega$.

The $\rho$ and $\omega$ contributions to the amplitude of Fig. 1(d) is:
\begin{eqnarray}
i{\cal M}_{fi}^\rho&=&-\frac{e\lambda_\rho}{\mu}
\frac{i\epsilon_{\mu\nu\alpha\beta} k^\nu q^\alpha
\varepsilon^\mu}{t-M_\rho^2}
\bar{U}_f \tau_3 \left
\{
g_v^\rho\gamma^\beta-\frac{g_t^\rho}{2M}i\sigma^{\beta\delta}(q-k)_\delta
\right \} U_i ,\\
i{\cal M}_{fi}^\omega&=&-\frac{e\lambda_\omega}{\mu}
\frac{i\epsilon_{\mu\nu\alpha\beta} k^\nu q^\alpha
\varepsilon^\mu}{t-M_\omega^2}
\bar{U}_f \left
\{
g_v^\omega\gamma^\beta-\frac{g_t^\omega}{2M}i\sigma^{\beta\delta}(q-k)_\delta
\right \} U_i.
\end{eqnarray}
The $\rho$ contributes to the isovector amplitudes, while
$\omega$ contributes to the isoscalar amplitudes,  Eq.(23).
 Due to near degeneracy of the $\rho$ and $\omega$ masses, one can
combine their amplitude for the proton target into one effective
coupling with the following set of coupling constants:
\begin{eqnarray}
\lambda_\rho g_v^\rho+\lambda_\omega g_v^\omega&=&5.93\pm 0.82, \\
\lambda_\rho g_t^\rho+\lambda_\omega g_t^\omega&=&17.50\pm 2.57.
\end{eqnarray}
We note here a remark made by Bernard {\em et al.} \cite{bernard1},
who point out that the tensor coupling is absent if chiral symmetry is insisted
upon.
However, in our effective Lagrangian approach, this coupling enters to mimic
chiral symmetry violation. This is true also of other effective
theories\cite{ren}.

Expanding the matrix elements in Eqs.(46), (47) in terms of the invariant
amplitudes $M_j$ of Eq.(19) yields
\begin{eqnarray}
A_1^V&=&\frac{e\lambda_V}{\mu}\frac{g_t^V}{2M}\frac{t}{t-M_V^2},\\
A_2^V&=&-\frac{e\lambda_V}{\mu}\frac{g_t^V}{2M}\frac{1}{t-M_V^2},\\
A_3^V&=&0,\\
A_4^V&=&-\frac{e\lambda_V}{\mu}g_v^V\frac{1}{t-M_V^2},
\end{eqnarray}
for the proton target with $V$ being either $\rho$ or $\omega$. For the
the neutron target the $\rho$ amplitudes are to be multiplied by $-1$.

An important theoretical issue here is the r\^{o}le of a form factor at the
$VNN$ vertex.
Brown and coworkers\cite{brown} have used a form factor of the type
\begin{equation}
F(t)=\frac{\Lambda^2-M_V^2}{\Lambda^2-t},
\end{equation}
preferring a value of $\Lambda^2\sim 2 M_V^2$. The nucleon-nucleon
interaction studies\cite{gross,machleidt} prefer a value of
$\Lambda^2\simeq 2.0 GeV^2$. Both
values are used in this work, and fixed during the fitting procedure.
The gauge invariance is preserved,
when a form factor is introduced at the VNN vertex, since the photon
couples through $F^{\mu\nu}$, as shown in Eq.(43).

\subsection{Nucleon resonance excitation}

The $s$- and $u$- channel resonance exchanges [Fig. 1(e)-1(f)] complete
the tree-level amplitude for the process (1). Here there are two
simplifications. First, since the eta meson has zero isospin, only
isospin $1/2$ nucleon resonances are allowed. Second, below $2$ $GeV$ c.m.
energy, only two nucleon resonances have significant decay branching
ratio into the $\eta N$ channel\cite{feltesse,baker,pdg}:
these are the $N^{*}(1535)$ and $N^{*}(1710)$ resonances,
with the respective $\eta N$ branching ratios being $50\%$ and $20-40\%$.
The other three resonances that were considered in this work are
$N^{*}(1440)$, $N^{*}(1520)$, $N^{*}(1650)$. The last two have $0.1\%$
and  $1\%$
as respective branching ratios into the $\eta N$ channel. The Roper resonance
$N^{*}(1440)$ lies below the $\eta N$ threshold of $W=1487 MeV$ and can
couple to the eta channel only due its large width. Its $\eta N$
coupling is unknown\cite{pdg}. Since the region
around c.m. energy $1535MeV$ is the primary interest of this work, other
resonances will have small contribution, mainly because they have large
mass and small coupling to the $\eta N$ channel. Therefore, in the
present investigation, the resonances included are
$N^{*}(1440)$, $N^{*}(1520)$, $N^{*}(1535)$, $N^{*}(1650)$, and
$N^{*}(1710)$. A
summary of some of the properties of these resonances as given
by Particle Data Group in the 1990 edition\cite{pdg}
is shown in Table \ref{table2}.
The more recent values as given in the 1992 edition\cite{pdg}
are also considered here.
They are  also summarized in  Table \ref{table2}.  The masses and
widths of the resonances considered here are more or less the same in both
editions, with
the following exceptions: (1) The  nominal value of the decay
width of $N^{*}(1440)$ is now $350MeV$. (2) The branching ratio for the process
$N^{*}(1710) \rightarrow \eta N$ is not very precisely known: it can vary
between $20$ to $40\%$.

\subsubsection{Spin-$1/2$ resonances}

\newcommand{\nstar}{\mbox{$N^{\textstyle \star}$}}

For a spin-$1/2$ nucleon resonance the vertex factors are similar to
those of the nucleon Born terms discussed earlier.
However, it differs at the $\gamma N \nstar$ vertex
in  that the coupling of the type given by the first term on the
r.h.s of Eq. (31) is absent, since its presence violates gauge invariance,
given  the inequality of the masses of the $\nstar$ and $N$.
Analogous to Eq.(29) and
(31), the spin-$1/2$ interaction Lagrangians are
\begin{eqnarray}
{\cal L}^{PS}_{\eta N R}&=&-ig_{\eta N R}
\bar{N} \Gamma R \eta + H.c., \\
{\cal L}^{PV}_{\eta N R}&=&\frac{f_{\eta N R}}{\mu}
\bar{N} \Gamma_{\mu} R \partial^{\mu}\eta + H.c., \\
{\cal L}_{\gamma NR}& = &{e \over{2(M_R+M)}} \bar{R}
(k^s_R+k^v_R\tau_3)\Gamma_{\mu \nu}NF^{\mu \nu}+H.c.,
\end{eqnarray}
where R is the generic notation for the resonance, $M_R$ its mass.
The transition magnetic couplings for the proton and neutron targets
are respectively $k^p_R=k^s_R+k^v_R \; and \; k^n_R=k^s_R-k^v_R $.
The operator structure for the $\Gamma$,$\Gamma_\mu$ and
$\Gamma_{\mu \nu}$ are :
\begin{eqnarray}
\Gamma = 1,& \Gamma_\mu=\gamma_\mu, &
\Gamma_{\mu \nu}=\gamma_5 \sigma_{\mu \nu}, \\
\Gamma =\gamma_5,&\Gamma_\mu=\gamma_\mu\gamma_5, &
\Gamma_{\mu \nu}=\sigma_{\mu \nu},
\end{eqnarray}
where (58) and (59) correspond to nucleon resonances of odd and
even parities respectively. Note that different parity states are
accounted for by inserting $\gamma_5$ matrix in the appropriate place.
In Eqs.(55), (56), the ambiguity in the meson-nucleon-resonance coupling
calls for the pseudoscalar ($PS$) or the pseudovector ($PV$) options. As
 shown in section III.A, the $PV$ coupling could also be introduced
through a unitary transformation of the nucleon field (Eq.(33))
to first order in the strong coupling. In the present case one needs to
redefine not only the nucleon field but also the resonance field in the
total $PS$ Lagrangian
(${\cal L}_{Free}+{\cal L}^{PS}_{\eta N R}+{\cal L}_{\gamma NR}$). The
appropriate transformations are
\begin{eqnarray}
&&N\rightarrow N+i\frac{g_{\eta N R}}{(M_R\pm M)}\Gamma_{\eta R}, \\
&&R\rightarrow R+i\frac{g_{\eta N R}}{(M_R\pm M)}\Gamma_{\eta N},
\end{eqnarray}
which  give
\begin{equation}
{\cal L}^{PV}={\cal L}^{PS} \pm i\frac{ef_{\eta N R}}{(M_R+M)\mu}
\bar{N}(k^s_R+k^v_R \tau_3)\sigma_{\mu\nu}\gamma_5 NF^{\mu\nu} \eta,
\end{equation}
the upper sign corresponding to even parity resonance.
$\Gamma$ is given by Eqs.(58), (59) and
\begin{equation}
\frac{f_{\eta N R}}{\mu}=\frac{g_{\eta N R}}{(M_R\pm M)}.
\end{equation}
To first order in $eg_{\eta N R}$,
the
$\gamma\;N\rightarrow\;R\rightarrow\;\eta\;N$
matrix elements for odd parity resonance in the $PS$ coupling are
\begin{eqnarray}
i{\cal M}_{fi}^{PS,R}&=&\frac{eg_{\eta N R}}{(M+M_R)} \bar{U}_f
\{(k^s_R+k^v_R\tau_3)\}
\left [\frac{\gamma\cdot (p_i+k)+M_R}{s-M_R^2}
\gamma_5\gamma\cdot k\gamma\cdot\varepsilon \right.
\nonumber \\
&&\left. + \gamma_5\gamma\cdot k\gamma\cdot\varepsilon
\frac{\gamma\cdot (p_f-k)+M_R}{u-M_R^2} \right ] U_i.
\end{eqnarray}
For the $PV$ case one can either start from the $PV$ Lagrangian or
add the equivalence breaking term given by Eq.(35), with the appropriate
couplings, to $PS$ matrix elements. The matrix elements of the even
parity resonances are deduced from the odd parity ones with the
following correspondences
\begin{equation}
i{\cal M}_{fi}^{R}(J^\pi=1/2^+)=
-i{\cal M}_{fi}^{R}(J^\pi=1/2^-,M_R \rightarrow -M_R  ),
\end{equation}
leaving the term $(M+M_R)$ in the denominator intact.
This can be easily checked by considering the nucleon as the
intermediate state and comparing the result with Eq.(32). The amplitudes
resulting from the spin-$1/2$ resonances in the $PS$ coupling are:
\begin{eqnarray}
A_1^R&=&\pm \frac{eg_{\eta N R}k_R}{(M+M_R)}
(M\pm M_R)\left [\frac{1}{s-M_R^2}+\frac{1}{u-M_R^2} \right ], \\
A_2^R&=&0,\\
A_3^R&=&\pm \frac{eg_{\eta N R}k_R}{(M+M_R)}
\left [\frac{1}{s-M_R^2}-\frac{1}{u-M_R^2} \right ], \\
A_4^R&=&\pm \frac{eg_{\eta N R}k_R}{(M+M_R)}
\left [\frac{1}{s-M_R^2}+\frac{1}{u-M_R^2} \right ],
\end{eqnarray}
with $+(-)$ sign corresponding to negative (positive) excited state and
$k_R=k^s_R\pm k^v_R$. The amplitudes due to the $PV$ coupling can be
obtained as indicated in Eqs.(40), (41) with the appropriate coupling.

\subsubsection{Spin-$3/2$ resonances}

The spin-$3/2$  excited state involved in this energy region is the odd
parity isospin-$1/2$ $N^{*}(1520)$ resonance.
Even though its $\eta N$ coupling
(see Tables \ref{table2}, \ref{table3}) is very small, its off-shell effects
could be
important. We have here important differences with the recent work of
Tiator {\em et al.} \cite{tiator}, who have ignored these effects.
We discuss the spin-$3/2$ propagator problem in the Appendix D.

In studying pion-nucleon  scattering and pion photoproduction from
nucleon at low energies, it is necessary to include, in the theoretical
calculations, the  contribution of the $\Delta(1232)$ resonance.
The question then arises as to how to treat the $\gamma N \Delta$ and $\pi
N \Delta$ vertices, which, in the usual approach contain off-shell terms.
These interactions have been discussed extensively in
the literature\cite{bdm}. In the
present case, the strong and electromagnetic three-point functions are
constructed in analogy with those of the
$\Delta(1232)$ resonance\cite{dmw,olsson,gourdin}, apart from taking
care of the isospin factors and the odd parity of the $N^{*}(1520)$. These are
\begin{eqnarray}
{\cal L}_{\eta NR}&=&\frac{f_{\eta NR}}{\mu} \bar{R}^\mu
\theta_{\mu\nu}(Z)\gamma_5 N \partial^\nu \eta + H.c., \\
{\cal L}^1_{\gamma NR}&=&\frac{ie}{2M}\bar{R}^\mu \theta_{\mu\nu}(Y)
\gamma_{\lambda}(k^{s\;(1)}_R+k^{v\;(1)}_R\tau_3)NF^{\lambda\nu} + H.c., \\
{\cal L}^2_{\gamma NR}&=&-\frac{e}{4M^2}\bar{R}^\mu\theta_{\mu
\nu}(X)
(k^{s\;(2)}_R+k^{v\;(2)}_R\tau_3)(\partial_\lambda N)F^{\nu\lambda} + H.c.
\end{eqnarray}
Here, $R^\mu$ is, for example,  the $N^{*}(1520)$ vector-spinor, and
\begin{equation}
\theta_{\mu \nu}(V)=g_{\mu
\nu}+[\frac{1}{2}(1+4V)A+V]\gamma_\mu\gamma_\nu;\;\;\;\;\; V=X,Y,Z.
\end{equation}
Here $A$ is an arbitrary parameter defining the so-called point transformation
(see Appendix, Eq.(D4)).
The interaction Lagrangians above have been constructed in such a way that they
are invariant under the same point transformation as the free one. The
form of $\theta_{\mu \nu}$ gives the most general interaction,
limited to the number of derivatives appearing in the Lagrangians in
Eqs.(70), (71) and (72)
 preserving the symmetry of the free Lagrangian. As a result, the
physical scattering amplitudes will be independent of $A$, according to a
theorem proved by Kamefuchi,O'Raifeartaigh and Salam\cite{kamefuchi}.
We  choose $A=-1$ for algbraic simplicity. The
parameters $X,Y,Z$, often referred as off-shell parameters,
are arbitrary, and will appear in the physical amplitudes. There
have been many theoretical attempts\cite{peccei,nath} to fix these parameters
but none have
been successful. There have also been some claims\cite{nath}, based on field
theoretic
arguments originally formulated by Fierz and Pauli\cite{fierz},
that the second coupling term [Eq.(72)] should be absent
and a special choice of values for
the parameters $Z$ and $Y$ is required. The choice is
$ Z=\frac{1}{2},Y=0,k^{(2)}_R=0$. One curious consequence of this,
arising from the absence of the gauge coupling $k^{(2)}_R$, is that the
dynamical freedom of two independent electromagnetic multipoles at the
$\gamma N N^{*}$ vertex is lost. Thus, the magnetic quadrupole ($M2$)
to the electric dipole ($E1$) amplitude ratio for the $N^{*}(1520)$ radiative
decay is fixed kinematically, rather than dynamically, yielding the ratio
 $M2/E1=(M_R-M)/(3M_R+M)\simeq 11\%$.
Available analyses\cite{pdg} give $31\% \leq M2/E1 \leq 56 \%$,
inconsistent with $k^{(2)}_R=0$ for the transition. Similar results
would follow for the $\Delta(1232)$ and other spin-$3/2$ baryons and are
discussed in the Ref.\cite{bdm} along with the various choices for the
off-shell parameters available in the literature.
The approach followed by the Refs.\cite{dmw,olsson} to fit these
off-shell parameters to the data will be adopted here.

The scattering amplitudes for the spin-$3/2$ odd parity resonance excitation in
the
$s$-channel are
\begin{eqnarray}
i{\cal M}_{fi}^{(1),R}&=&C_1 \bar{U}_f
q^\mu \theta_{\mu\nu}(Z)\gamma_5 P^{\nu\lambda}(p)
\theta_{\lambda\sigma}(Y)
(\gamma\cdot\varepsilon k^\sigma-\varepsilon^\sigma \gamma\cdot k) U_i, \\
i{\cal M}_{fi}^{(2),R}&=&C_2
\bar{U}_f ^\mu \theta_{\mu\nu}(Z)\gamma_5 P^{\nu\lambda}(p)
\theta_{\lambda\sigma}(X)
(p_i\cdot k \varepsilon^\sigma -p_i\cdot \varepsilon k^\sigma) U_i,\\
&& C_1=-\frac{e k_R^{(1)} f_{\eta NR}}{2M\mu},\;\;\;
C_2=-\frac{e k_R^{(2)} f_{\eta NR}}{4M^2\mu}, \nonumber
\end{eqnarray}
where $k_R^{(i)}=k_R^{s,(i)}\pm k_R^{v,(i)}$, with $+(-)$ corresponding
to the proton (neutron) target. As a check, the even
parity spin-$3/2$ exchange (such as $\Delta(1232)$, apart from isospin)
scattering amplitudes are obtained by noticing that
\begin{equation}
\gamma_5 P_{\nu\lambda}(p,M_R)=-P_{\nu\lambda}(p,-M_R)\gamma_5.
\end{equation}
Therefore,
\begin{eqnarray}
i{\cal M}_{fi}^{(1),R}(3/2^+)&=&-i{\cal M}_{fi}^{(1),R}(3/2^-,
M_R\rightarrow -M_R), \\
i{\cal M}_{fi}^{(2),R}(3/2^+)&=&i{\cal M}_{fi}^{(2),R}(3/2^-,
M_R\rightarrow -M_R).
\end{eqnarray}
For the $u$-channel, the strong and electromagnetic vertices
need to be interchanged.
The expressions for the invariant amplitudes are lengthy and
are collected in the Appendix E
for convenience.

\subsubsection{Resonance couplings}

The two independent coupling constants of each spin-$1/2$ resonance can
be combined into one effective constant $k_R^{p}g_{\eta N R}$. The
three independent couplings of the spin-$3/2$ resonance can be
grouped into two effective coupling constants $k_R^{p\;(1)}f_{\eta N R}$ and
$k_R^{p\;(2)}f_{\eta N R}$, representing the two independent
interactions at the electromagnetic vertex. It is more useful to express
these couplings in terms of the experimentally observable quantities such
as partial decay widths. For illustration, the dominant resonance
$N^{*}(1535)$ case will be discussed in detail, and a similar
procedure can be applied to the remaining resonances.

 From the Lagrangian density (57), the S-matrix element for the process
$N^{*}(1535)\rightarrow \gamma p$ may be written as
\begin{equation}
S_{fi}=-i\sqrt{\frac{M}{2kE_p}}
\frac{\delta^4(p_p+k-p_R)}{(2\pi)^{1/2}}{\cal M}_{fi},
\end{equation}
where
\begin{equation}
{\cal M}_{fi}=\frac{ek_R^p}{(M_R+M)} \bar{U}_p\gamma\cdot k
\gamma\cdot \varepsilon \gamma_5 U_R.
\end{equation}
Here $E_p=\sqrt{M^2+k^2}$ and $k$ are the energies of the proton and the
photon respectively defined in the frame where $\vec{p}_R=0$. Upon
integrating over the phase space and averaging over the initial spin and
summing over the final spins, one obtains the radiative width
\begin{eqnarray}
\Gamma_{N^{*}(1535)\rightarrow \gamma p}
&=& \left (\frac{ek_R^p}{M_R+M} \right )^2 \frac{k^2}{2\pi}
\frac{(M_R^2-M^2)}{M_R}.
\end{eqnarray}
Alternatively, it may be expressed in terms of the more familiar helicity
amplitude $A_{1/2}^p$\cite{copley} through
\begin{equation}
\Gamma_{N^{*}(1535)\rightarrow \gamma p}=
\frac{k^2}{\pi}\frac{M}{M_R} |A_{1/2}^p|^2.
\end{equation}
Comparing the two expressions, one can easily deduce
\begin{equation}
|A_{1/2}^p|^2= \left (\frac{ek_R^p}{M_R+M} \right )^2
\frac{(M_R^2-M^2)}{2M}.
\end{equation}

For the strong coupling, one can use either (55)
or (56). The two Lagrangian densities lead to the same result, as
they should for an on-shell resonance state. The S-matrix element is
\begin{equation}
S_{fi}=-i\sqrt{\frac{M}{2\omega E_N}}
\frac{\delta^4(p_N+q-p_R)}{(2\pi)^{1/2}}{\cal M}_{fi},
\end{equation}
with
\begin{eqnarray}
{\cal M}_{fi}
&=&-i\frac{f_{\eta N R}}{\mu} \bar{U}_N\gamma\cdot q U_R,\nonumber \\
&=&-i\frac{f_{\eta N R}}{\mu} (M_R-M) \bar{U}_N U_R=-ig_{\eta N
R}\bar{U}_N U_R.
\end{eqnarray}
Here $E_N=\sqrt{M^2+q^2}$, and $\omega$ and $q$ being the energy and
momentum of the $\eta$ meson. The $\eta N$ partial decay width is given
by
\begin{equation}
\Gamma_{N^{*}(1535)\rightarrow \eta N}=\frac{g^2_{\eta N R}}{4\pi}
\frac{q(E_N+M)}{M_R}|_{W=M_{R}}.
\end{equation}

At this stage, it is important to show the relationship between the
resonance couplings and the experimentally extracted multipoles or
helicity elements describing the process
$\gamma + N \rightarrow N^{\textstyle \star} \rightarrow \eta + N$.
These relations are already known for pion photoproduction.
Signs arising from the $N^{\textstyle \star}\rightarrow
\pi + N$ decay are involved .
Following the prescription given by the Particle Data Group (PDG-76)\cite{pdg},
\begin{eqnarray}
A_{\ell \pm}&=&\mp S_X C_{X N}^I A_{1/2}, \\
B_{\ell \pm}&=&\pm S_X
\left ( \frac{16}{(2J-1)(2J+3)} \right )^{1/2} C_{X N}^I A_{3/2}, \\
C_{\ell \pm}&=&\mp S_X C_{X N}^I S_{1/2}.
\end{eqnarray}
$S_X$ describes the decay of the resonance into $X N$, where $X$ is
either $\pi$ or $\eta$, and
is given by
\begin{equation}
S_X= \left (\frac{1}{(2J+1)\pi}\frac{k_R}{q^X_R}\frac{M}{M_R}
\frac{\Gamma_X}{\Gamma^2} \right )^{1/2}.
\end{equation}
Here $k_R,q^X_R$ are the photon and meson momenta respectively
in the c.m. frame and
evaluated at $W=M_R$ and $k^2=0$. $C_{X N}^I$ is a Clebsch-Gordan
coefficient related to the isospin coupling in the outgoing channel.
As an example, let us consider the amplitudes for
the $\pi^\circ p$ and $\eta p$ production via the $N^{*}(1535)$ resonance:
\begin{eqnarray}
A_{0+}^{\pi^\circ p}&=&-S_\pi \left (-\sqrt{\frac{1}{3}} \right ) A_{1/2}=
+\sqrt{\frac{1}{3}} S_\pi A_{1/2}, \\
A_{0+}^{\eta p}&=& -S_\eta A_{1/2}.
\end{eqnarray}

It is clear that the $A_{0+}^{\pi^\circ p}$ has the same sign as
$A_{1/2}$ and according to Table \ref{table3},  it is positive, consistent
with Walker's analysis\cite{walker}. Now, since the
relative sign between the $\pi$ and $\eta$ strong vertices is positive
(see Table \ref{table3}), and the electromagnetic vertices are the same,
one then expects that the $A_{0+}^{\eta p}$ is also positive. Therefore,
the sign appearing in Eq.(92) is misleading. One has to take into
account the isospin convention used and the relative
sign between the couplings of $N^{\textstyle \star}\rightarrow \eta N$
and $N^{\textstyle \star}\rightarrow \pi N$. Table \ref{table3}
gives the helicity
amplitudes and the partial decay width to $\eta N$ including the
relative sign to $\pi N$, as estimated by $PDG92$\cite{pdg},
for the resonances considered in this
work. The
corresponding quantities are also given in the constituent quark model
calculations of Koniuk and Isgur\cite{koniuk}, as well as  Capstick and
Roberts\cite{capstick}. There is  agreement in sign between the quark model
approaches and the PDG for the $N^{*}(1520), N^{*}(1535)$ and $N^{*}(1650)$
resonances.
Being below $\eta N$ threshold, the Roper resonance $N^{*}(1440)$ $\eta N$
coupling is not
very well-determined, but the quark model calculation\cite{capstick} indicates
a small
coupling with positive relative sign to the $\pi N$ coupling. The sign for
the $\gamma p\rightarrow N^{*}(1710) \rightarrow \eta p$ amplitude is
not determined from phenomenological studies\cite{baker},
but the quark model calculations\cite{capstick}  prefer a negative sign.
Therefore, the sign for this amplitude will be allowed to change during
the fitting procedure.

\subsubsection{Approximate unitarization of resonant amplitudes}

We now give an approximate unitarization procedure for the resonance excitation
amplitudes. We assume two-channel K-matrix, where the channels are
$I N\rightarrow N^* \rightarrow J N$ with $I,J = \pi,\; \eta$. This yields,
using the PS coupling of  meson-nucleon resonance,
the expression for the amplitude to excite $N^*(1535)$:
\begin{equation}
E_{0^+}^{R,PS}=-{{eg_{\eta NR}k^p} \over {8\pi W(M_R+M)}}
{{ab_\eta (W-M)(W+M_R)} \over {W^2-M^2_R+iM_R \Gamma_T(W)}},
\end{equation}
with
\begin{equation}
b_i^2(W)={{(W+M)^2-m^2_i} \over {2W}}, \;\;
a^2={{(W+M)^2} \over {2W}},
\end{equation}
\begin{equation}
\Gamma_T(W)=({{W+M_R}\over{2W}})\sum_{i}\left(
{{b^2_i(W)} \over {b^2_i(M_R)}}{{q_i} \over {q^R_i}}\Gamma_i
\right), \;\;\;\;i=\pi,\eta.
\end{equation}
$\Gamma_i$ is the partial decay width of $S_{11}$ into (iN), R being
$S_{11}$ here, $q_i's$ are the cm momenta of the meson i.
Again, we have for $S_{11}$,
\begin{equation}
|g_{\eta NR}|=\left({{4\pi M_R} \over {q^R_{\eta}b^2_{\eta}(M_R)}}\Gamma
_{\eta}\right)^{1/2}.
\end{equation}
Similar expressions for other resonances can be given.

\section{Results and discussion}
\subsection{Fitting Strategy}
Previous attempts\cite{homma,hicks,tabakin}
at the analysis of eta photoproduction data,before the work of Benmerrouche
and Mukhopadhyay\cite{bm1}, have not only suffered from the crudeness of the
data,
but also from {\em the lack of enough theoretical constraints} in restricting
the number of parameters fitted, $24$ or more. The effective Lagrangian
approach provides us
with a tremendous reduction in the
number of free parameters, eight in our case.
These are, $g_\eta$, four parameters associated with the spin-$1/2$
resonances and three parameters
associated with $N^{*}(1520)$.
The resonance masses and widths are taken from
analyses of strong interaction processes in order to reduce the number of
parameters. Four sets of resonance parameters are
used and are given in Table \ref{table4}.
 $KOCH$ and $CUTK$ parameters
are determined through the analyses of $\pi N$ scattering by
Koch\cite{koch} and Cutkosky\cite{cutkosky} respectively.
They differ drastically on the widths of the Roper and $N^{*}(1535)$
resonances. $BAKR$ parameters are taken from the
analysis of $\pi^- p \rightarrow
\eta n$ by Baker {\em et al.} \cite{baker}.
Earlier analysis\cite{feltesse}
of the same reaction did not include the whole data set;  therefore,
parameters from the Ref.\cite{feltesse} will not be considered here. $PDG92$
refers to
the nominal values estimated by the Particle Data
Group in the 1992 edition\cite{pdg}. The off-shell parameters
$\alpha,\beta,\delta$
associated with the
spin-$3/2$ field are not well established theoretically.
 The parameter $\alpha$, which appears at the strong interaction
vertex, should,
in principle, be determined from strong interaction processes such as
$\pi^- p \rightarrow \eta n$. From pion photoproduction analysis in the
$\Delta(1232)$\cite{dmw}, there are some indication that
$\alpha$ should be between zero and two,
but all off-shell parameters, in general, are not known.
In the present analysis, $\alpha$
will be varied from $-1$ to $3$ and the other two parameters ($\beta,\delta$)
will be determined from the fit to the data. Also, varying the $g_p^{(2)}$
electromagnetic coupling of the $D_{13}$ has not improved the fit
considerably and therefore the ratio $g_p^{(2)}/g_p^{(1)}$ is fixed to
the $PDG$ value of $0.69$. An improved version of the $CERN$
fitter routine, $MINUIT$ is used to minimize the weighted least-squares
function $\chi^2$
\begin{equation}
\chi^2=\sum_i \frac{(X_i-Y_i(a_1,\cdots,a_n))^2}{\sigma_{X_i}^2},
\end{equation}
where $X_i$ represent the experimental observables,
$\sigma_{X_i}$ are their standard
deviations, and $Y_i(a_1,\cdots,a_n)$ are the theoretical predictions
with $a's$ being the parameters of the theory. In the present case, the
observables are the differential cross section and the recoil nucleon
polarization, and the summation is over the energies and angles.
\subsection{Analysis of the older  data base}
The older  data base for the
differential cross section\cite{delcourt}-\cite{ukai}
for $E_\gamma$ between $725 MeV$ and $1200 MeV$ contains $137$ data points
(including the
recent measurement by Homma {\em et al.} \cite{homma}). The old polarization
measurements\cite{heusch} ($7$ data points, of which $5$ are at
$90^\circ$) for $E_\gamma$ between $725$ and
$1100~MeV$ are also included in the fit.
We use this data base to generate the first series of fits, called Fit A.
Results of this fit are summarized in Tables \ref{table5}  to \ref{table9}.
Note the effect of the off-shell parameter at the strong vertex:
for a given set of inputs for resonances, the effect of this parameter is not
large.
There is also some influence of the vector meson form factor through its
cut-off
parameter on the fitted results.

Below we discuss in detail our analysis of the old data base. Our comments
on the new data will suppliment these observations.

\subsubsection{Resonance characteristics}

In Tables \ref{table5} to \ref{table11}, all but the last two relevant to the
old data base, the resonances
 have been renamed  for brevity:
for example, $S_1$ denotes the $N^{*}(1535)$, $P_1$ the Roper resonance,
etc. The total $\chi^2$ is given separately for the differential
cross section ($XS$) and the recoil nucleon polarization ($POL$). The
total $\chi^2$ per degree of freedom
 is denoted by $\chi^2_{TOT}/DF$.
$\sqrt{\Gamma_\eta}A_{1/2,3/2}$ has been multiplied by $1000$ and the
helicity amplitudes $A_{1/2,3/2}$ are expressed in the standard units of
$10^{-3} GeV^{-1/2}$. As there are large differences in the masses and
widths of the resonances in various analyses, we have investigated
 the results of our fitting here how are affected by
a particular  choice of those parameters. One important funding is that
the quantity $\xi$,
characteristic of the photoexcitation of the $N^{*}(1535)$ resonance and
its decay into the $\eta$-nucleon channel, defined as
\begin{equation}
\xi=\sqrt{\chi'\,\Gamma_\eta}\,A_{1/2}/\Gamma_T,
\end{equation}
where $\chi'=Mk/qM_R$, $k$ and $q$ to be evaluated at $W=M_R$,
{\em is not} sensitive to uncertainties of the resonance parameters
and other  details of the effective Lagrangian approach. Taking a
simple average over the new results from PDG92
fits gives a rather precise determination of $\xi$\cite{note3}
\begin{equation}
\xi=(2.2\pm 0.2) \times 10^{-1} GeV^{-1}.
\end{equation}
This quantity should be of fundamental interest to a precision test of
hadron models.
Another important feature of this analysis is that the product
$[\sqrt{\Gamma_\eta} A_{1/2}]_{S1}$ extracted from the data
 does not depend on the
details of the background (given a set of resonances). Given a particular
choice of resonance parameters, indicated in parenthesis,
we find the following parameters for the $N^{*}(1535)$, using the old data
base:
\begin{eqnarray}
M_R (MeV),\Gamma_T(MeV)
&A_{1/2}(10^{-3} GeV^{-1/2}) \nonumber \\
1535,150(PDG92) & 97\pm 7 \\
1526,120(KOCH)
&  87\pm  6 \\
1517,180(BAKR)
& 104\pm 6 \\
1550,240(CUTK)
& 173\pm 9
\end{eqnarray}

Here the agreement between the first three sets of numbers is reasonable,
while the figures for the CUTK do not agree with the previous  ones.
Given the large difference in the  total width for the $N^{*}(1535)$
between the CUTK set and the others, this is not surprising. However, the fit
with $BAKR$ falls short of the experimental data (Figs. 4 and 5). This may be
due to unusually low value of the $\eta N$ branching ratio ($36\%$).
We refer the reader to a recent analysis by
Manley and Saleski\cite{manley}, who have extracted resonance
parameters using the isobar model to analyse the data for the
 $\pi N \rightarrow \pi \pi N$. Their inferred mass and width of
the $N^{*}(1535)$ agree well with the $PDG92$ nominal values.

We thus obtain, from the old data base, and using inputs from the $PDG92$,
\begin{eqnarray}
A_{1/2}=(97\pm 7) \times 10^{-3} GeV^{-1/2},
\end{eqnarray}
for the proton. This  lies in
between the predicted extremes of recent theoretical estimates in the
quark model\cite{koniuk,feynman,close,warns} ranging from $54$ to $162$,
in the same units. The latest of these  is from
Capstick\cite{capstick}, who has obtained a value  $A_{1/2}=76 \times 10^{-3}
GeV^{-1/2}$.
The corresponding value, reported by the $PDG(1992)$, extracted
from the pion photoproduction  data is $74\pm 11$\cite{pdg}.
The origin of the disagreement between the resonant amplitudes
extracted from eta and pion photoproduction data is not understood
at present. The quark model estimates are still too crude to be definitive
in testing the model. The issues of the trucation of the model
space and the lack of current conservation\cite{bmquark} are just some of
many unresolved issues.  Much work remain to be done here.

As far as the helicity amplitudes of the other resonances are concerned, the
old data set
does not permit an accurate extraction, except to allow the conclusion
that they are consistent, in general, with  the results extracted from
pion photoproduction analyses. The results obtained for each resonance
will now be briefly discussed.
\begin{enumerate}
\item
Assuming a branching ratio to $\eta N$ of $1\%$,
the $N^{*}(1650)$ photocoupling can be as  large as four times that
 obtained by pion photoproduction analysis and by the quark model
calculations. This might be due to the ambiguity in determining the
$\eta N$ branching ratio.
\item
The $N^{*}(1440)$ coupling is found to be very small indicating that
this resonance may not be a significant player in the ($\gamma,\eta$)
process. Recently, there have been some speculation\cite{carlson} that the
$N^{*}(1440)$ might be a candidate for the lightest hybrid state,
consisting of three valence quarks and one valence gluon. A precise
determination of the $N^{*}(1440)$ photocoupling can provide a
powerful tool to distinguish between the different internal structures
for this hadron: $q^3$ and $q^3 G$, where $q$ is a valence
quark and $G$ is a valence gluon.
\item
The $N^{*}(1520)$ off-shell contribution is found to be very
important and correlates with the nucleon and the vector meson
contributions.
However, this is not so in the case of $CUTK$, as can be seen by comparing
column 3 and 5 of Table \ref{table8}  and Table \ref{table9}.
This conclusion may be connected
with the relatively large width of the $N^{*}(1535)$ in the CUTK set.
 The off-shell
parameters
depend on the cutoff in the vector meson form factor and the choice of
coupling for the $\eta NN$ vertex (compare for example Tables \ref{table6}
and \ref{table7}).
The photocouplings of the $N^{*}(1520)$ are consistent with pion
photoproduction studies.
\item
The $N^{*}(1710)$  is poorly determined from the
$\gamma N \rightarrow \pi N$ reactions.
{\em This resonance photocoupling could be well-determined from the
present analysis if there were enough data around $W \simeq 1710$ $GeV$}.
All it can
be said is that our fits favor a positive sign for the
product $[\sqrt{\Gamma_\eta} A_{1/2}]_{P2}$, in {\em disagreement} with the
quark model prediction of Koniuk and Isgur but {\em in agreement}
with  the Capstick and Roberts result (see Table \ref{table3}).
\end{enumerate}

\subsubsection{Measured quantities in the experiments on eta photoproduction}

\paragraph{Differential cross-section}

In general, reasonable fits to the available data  on
differential cross section are obtained.
The best fit to the data is obtained with
the $PDG$ parameters. Sample fits using
$PDG$, $KOCH$ and $BAKR$ resonance parameters are displayed in
Figs. 2, 3, 4  for the c.m. angle of
$90^\circ$.
The $s$-channel excitation of the $N^{*}(1535)$
resonance dominates the differential cross section, while the
$u$-channel contribution is found to be negligible.
The three sets are in good agreement with the data, but start
deviating from each other above $E_\gamma=1100MeV$. Around this
energy region, the $N^{*}(1710)$ is the dominant resonance contribution
due to its large $\eta N$ branching ratio.

\paragraph{Total cross-section}

The old  data base on the total cross section\cite{baldini} suffer from poor
photon energy
resolution and counting statistics, and thus limit the quality of physics
extractable from them. The model prediction is in agreement with the
data for the three sets of resonance parameters as displayed in
Fig. 5. There is one data point well outside our fit.
The $BAKR$ parameters tend to unnderestimate the total cross-section.

\paragraph{Polarization observables}

Our predictions for polarized target asymmetry
and photon asymmetry are also shown.
All three sets show more or less the same behavior for the
polarization observables below the photon lab energy of one $GeV$,
but yield very different predictions above this energy.  Therefore,
polarization observables should provide a
more stringent test of the model. The  meagre data on the recoil
nucleon polarization are too poor to be of any quantitative value.

\subsection{Analysis of the Bates angular distribution data}

The recent eta photoproduction experiment  at the Bates Laboratory by the
Pittsburgh-Boston-LANL collaboration\cite{daehnick}
will now be discussed. This group has been able to measure the angular
distribution for
($\gamma,\eta$) reaction at photon lab energies of $729$ and $753 MeV$ at six
angles.
These data are more or less  flat as a function of angle (Fig. 6) at
$E_{\gamma}\, = \,729\,MeV$, consistent with the predictions of our effective
Lagrangian
approach, using the parameters of the Fit A. However, the data set at
$E_{\gamma}\,=\,753\,MeV$ exhibit a deviation from isotropy, in disagreement
with the prediction of the Fit A (Fig. 6). This suggests some {\em
inconsistency} between the new Bates data and the old data set from which the
Fit A
has been derived.

We now use the older data set and the Bates data together for a
global fit, Fit B. The resultant resonance parameters are shown in Table
\ref{table10}. This is a compromise fit between two somewhat incompatible data
sets.

Finally, we can, of course, use the Bates data alone
and try to fit the effective
Lagrangian parameters.  In so doing, we shall keep the $N^{*}(1535)$,
$N^{*}(1650)$ and $N^{*}(1710)$ parameters fixed at the Fit B level, along with
the $\eta NN$ coupling constant,
as this data set, by itself, cannot yield informations  on the
properties of all of these resonances, due to limited
energy coverage. Instead, we
use this data set to explore the nature of coupling of the nucleon
Born terms and the properties of the $N^{*}(1440)$ and $N^{*}(1520)$
resonances.
These fits, called Fit C, are shown in Fig. 7. The resultant fit still shows
dominance of the  $N^{*}(1535)$ in the differential cross-section.
The parameters for the resonances change somewhat (Table \ref{table11}),
compared with what we have obtained from the old data base.
As discussed below in the subsection 5,{\em the
$E_{0+}$ amplitude, extracted from the Bates data, is not in agreement with
those from the other data sets.}

\subsection{A look at the preliminary Mainz data}

The results of an exhaustive eta  photoproduction experiment\cite{krusche}
from the Mainz Microtron are available in a preliminary form, following
their presentations at the Trieste, Perugia and Dubna
conferences\cite{krusche}.
We emphasize the word {\em preliminary}, as these data are yet to be published
in a definitive form. The data, presented so far, are the angular
distributions at the photon laboratory energies of $E_{\gamma}\,=\,722.5,\,
737.5,
\,752.5,\,767.5$ and $782.5$ $MeV$, though the absolute normalizations of
these distributions are yet to be determined. Also available from the recent
Dubna
conference are the preliminary  Mainz data on the total eta
photoproduction cross-section,
again with the absolute normalization being arbitrary.  We compare them with
our predictions from the effective Lagrangian
approach, with parameters determined from the old world supply of data, Fit A.
The {\it shapes} of both the angular distributions and total cross-section
of the Mainz data are predicted rather nicely
(Figs. 8 and 9 respectively). The arbitrary
factor needed to bring the data of the differential
and total cross-section is the same. We are waiting with anticipation for the
definitive normalizations of these data.

We can get an idea of the importance of the preliminary Mainz  data on the
physics of the $N^{*}$ resonances, particularly $N^{*}(1535)$. The parameter
$\xi$, in units of $10^{-1}GeV^{-1}$, defined earlier,
extracted only from the Mainz data (Fit D) is
\begin{equation}
\xi = 2.0\pm 0.1,
\end{equation}
for the off-shell parameter $\alpha =-1$, and
\begin{equation}
\xi =2.3\pm 0.1,
\end{equation}
for the off-shell parameter $\alpha =+1$, for the vector meson
form factor cut-off parameter $\Lambda^{2}=1.2 GeV^{2}$. These compare
with the value
\begin{eqnarray}
\xi =2.2\pm 0.2
\end{eqnarray}
from our Fit A of the world's old data set. Thus, the error
on the $\xi $ parameter is somewhat reduced, using the Mainz data.
Taking $\Gamma_{\eta}=75 MeV$, the preliminary
Mainz  data yield a value of $A_{1/2}$, in units of
$ 10^{-3} GeV^{-1/2}$,
\begin{eqnarray}
A_{1/2}=88, & 101,
\end{eqnarray}
for the above two cases, consistent with our results from the older
data set, Eq.(100). It is substantially larger
than the latest results from the quark model\cite{capstick}
 $75\times
10^{-3} GeV^{-1/2}$, and that from the analysis of  pion photoproduction.
The results (Eqs. 101,102,104) are
 subject to revision, if the Mainz data
change substantially, as they are reanalyzed, but the basic conclusion of
their importance in determining the $\xi $ parameter should remain valid. For
comparison, we quote here the values of $\xi $ and $A_{1/2}$ extracted from
the Bates data under similar theoretical assumptions in analysis:
\begin{eqnarray}
\xi = 2.1\pm 0.1, & 1.8\pm 0.3, \nonumber \\
A_{1/2}= 91, & 78.
\end{eqnarray}

\subsection{The $E_{0^{+}}$ amplitude at threshold}

The threshold eta photoproduction and its impact on the determination of the
$\eta NN$ coupling constant is the last topic we wish to address here.
It is particularly interesting to contrast the $E_{0^{+}}$ amplitude for
the $\eta$ photoproduction with that for the $\pi^{0}$ photoproduction
at their respective thresholds. Table \ref{table12}  demonstrates
the real part of the $E_{0+}$ amplitude for the
$\eta$ photoproduction off the proton at threshold, contrasted with the same
for the $\pi^\circ$ photoproduction.
While there is clear preference for
the PV coupling at the $\pi^\circ NN$ vertex, this is not so for the $\eta$
meson, as we have pointed out earlier.
Here, we have given the $E_{0^{+}}$ amplitude as determined from the Fits A,
B, C and D, demonstrating the relevance of different data sets in the context
of the $E_{0^{+}}$ amplitude. {\it All fits agree in the dominant
role of the $N^{*}(1535)$ resonance}.

Interestingly, the value of the
$\eta NN$ coupling constant extracted from the fits is
only mildly sensitive to
the choice of the $PV$ or $PS$ coupling at the meson-nucleon vertex.
 From fits to all data sets, using the resonance parameter set of $PDG92$,
we get
\begin{equation}
0.2 \leq g_\eta \leq 6.2.
\end{equation}
For the $\eta$ photoproduction, the vector meson contributions are sizable,
but the $N^{*}(1535)$ excitation amplitude in the $s$-channel stands out, in
contrast to the $\pi^\circ$ case, where both the vector meson
and the $\Delta(1232)$
contributions are minor. Another important contribution to eta
photoproduction is  from the $N^{*}(1520)$,  with its off-shell
effects tending to interfere destructively with the the large contribution
of the $PS$ Born terms, or constructively with the smaller contribution of
the $PV$ Born terms.

\section{CONCLUSIONS }

Given the renewed  theoretical interest arising from the prospect of testing
$QCD$ in the non-perturbative domain by computing hadron properties, and
the experimental possibilities of exploring many of these properties in
the novel electron/photon facilities now under development, particularly
CEBAF, eta photoproduction
 on the proton have been investigated in the
$N^{*}(1535)$ resonance region, with a view to help understand
the structure of the nucleon and its excited states. In
this paper, the goal has been to extract
the product of the  electric dipole transition amplitude
$\gamma p\rightarrow N^{*}(1535) $ and the decay amplitude $N^{*}(1535)
\rightarrow \eta p$, from the
existing  old experiments and new ones. The dominant tree-level contributions
 considered here have been computed
in the framework of the effective Lagrangian formalism, proven to be very
successful in describing pion photoproduction in the $\Delta (1232)$
region. Unlike the $\pi^\circ$ case,
there is no compelling reason to choose the pseudovector($PV$)
 form of the $\eta NN$
(or $\eta NN^{\textstyle \star}$) coupling, and we have investigated
both the PV and the pseudoscalar(PS)
 couplings at the $\eta NN$ and the $\eta NN^{*}$
vertices. We have taken into account various  background contributions,
and have attampted
to extract information on the excitation and decay of the $N^{*}(1535)$
resonance. Our conclusions are as
follows: \\[1em]
1. Unlike the pion, where there is a
clear preference for the $PV$ coupling at the meson-nucleon vertex,
seen in the threshold $\pi^\circ$
photoproduction data on the proton, the present experimental data
on $eta$ photoproduction do not
distinguish between pseudovector and pseudoscalar couplings, as
contributions from various resonances and non-resonant background
compensate. This is not surprising, as the chiral symmetry is strongly
broken by the eta mass. Likewise, there is no strong preference for
either coupling at the $\eta NN^{*}$ vertices. \\[1em]
2. The extracted  amplitude
 shows that the $\pi^\circ$ and the $\eta$
photoproduction processes  near
threshold have very significant differences, even as they share the
common contributions, such as the nucleon Born terms, the basis
for the predictions of the low energy theorems (LET) for the pion
case. Among these differences, the contribution to the
$\eta$ photoproduction by the $s$- channel excitation of the
$N^{*}(1535)$ resonance is obvious. The situation is
quite different in the $\pi^\circ$ case, where one probes only the
nucleon Born terms, and, through it, the chiral symmetry breaking effects,
with minor contributions from the $\Delta(1232)$ excitation.
Thus, the chiral symmetry breaking effects are hard
to quantify in the $\eta$ case. \\[1em]
3. Many previous  attempts at the analysis of $\eta$ photoproduction data have
not only suffered from the crudeness of the data, but also from the lack
of enough theoretical constraints in restricting the number of
parameters fitted, {\em twenty four or more}.
The effective Lagrangian provides us
with a {\em tremendous reduction} in the number of free parameters,
{\em eight in the present work}.
 The data base is immensely improved with the addition of the Bates and
Mainz data sets, the latter still in their preliminary form.
 \\[1em]
4. The $E_{0+}$ amplitude at the eta photoproduction threshold,
inferred from the new Bates data, does not agree with those extracted from the
older data set and the new Mainz data. However, the
conclusions on the $N^{*}(1535)$ excitation amplitude are similar in analyses
of
all of these data sets. \\[1em]
5. Our  analysis yields a {\em precise estimate} of the
product $\Gamma_\eta^{1/2} A_{1/2}$ for the
$N^{*}(1535)$, which is quite insensitive to  the uncertainties in the other
resonance properties known  thus far.
For a given set of  these resonance
parameters, it is {\em not sensitive} to the detail of the background, such as
the off-shell parameters of $N^{*}(1520)$, the form factor in the vector
meson amplitude and the type of the meson-nucleon coupling.
\\[1em]
6. The present experimental situation on photoproduction at higher energies
(W$\geq$ 1400 MeV)
is not precise enough to extract any meaningful information about
the contributions from resonances other than the $N^{*}(1535)$.
\\[1em]

We should stress that
precise data on the polarization observables are missing and are badly
needed. These would be valuable to test the models for the background
contributions.

As to the future extension  of this work, we must mention the prospect
for a rigorous investigation of the unitarity effects.
The application of unitarity to the eta photoproduction process
is a very complicated task,
because there are a fair number of channels coupled to the process and many
of them contain more than two particles in the final state, as in the
case of multipion
production channels. Therefore, even a modest unitarization of the amplitude
should include at least
five channels. There is no consistent partial
wave analysis of  the channel, $\pi N \rightarrow \eta N$, one of the most
important; also, no experimental
information
on the $\eta N \rightarrow \eta N$ process is available.
A successful understanding of strong interaction
processes will help implement unitarity
 to electromagnetic processes.
Therefore, the pion-induced eta production, to be studied at
facilities like COSY in J\"{u}lich, Germany,
could help a great deal. These could be investigated using the
Lagrangians for the strong vertices considered here.
Valuable information on the
strong decays of the type $N^{\textstyle \star} \rightarrow N \eta$ could be
extracted. Existing treatments of these processes
  give acceptable representation of the data, but do not make any
precise connection with hadron models.

An obvious extension of the present analysis is the photoproduction of
$\eta^{\prime}$ meson on the nucleon, which is underway\cite{zmb}.
A comparative study of
$\eta$ and $\eta^{\prime}$ may lead to valuable information on how $\eta$
and $\eta^{\prime}$ interact with nucleons and its excited states.

It is also interesting to use the $\eta$ photoproduction
amplitude as an effective impulse  operator to
study photoproduction of eta mesons off nuclei. Such a
theoretical study  has been initiated by Doyle\cite{doyle},
and followed by others\cite{tiator}.
One would like to learn more about the properties of the $N^{*}(1535)$ in the
nuclear
medium\cite{carrasco}.
One important task is to use the
effective Lagrangian, developed here, to investigate the photoproduction
 of the
$\eta$ meson on the deuteron, where
there seems to remain a serious discrepancy between the recent theoretical
investigation\cite{halderson} and photoproduction experiments.
{\it This process is very crucial in extracting the electromagnetic transition
amplitude of $N^{*}(1535)$ on the neutron, for which hadron models have
clear-cut
predictions}.

Hopefully, this paper has provided a good motivation for future work
involving physics of eta mesons and excited baryons.
Careful studies of electromagnetic as well
as hadronic eta production processes are needed to obtain a more complete
picture of the $\eta N$ and $\eta$-nucleus interactions. New
facilities, such as  CEBAF and COSY, would be
good places to explore this subject further.
Particular mention should be made of
 the superior design  capabilities  of a
device called the
CEBAF Large Acceptance Spectrometer (CLAS):  its almost $4\pi$
solid angle coverage and the possible use of polarized targets and beams
in conjunction with it. Our  work lays the basic foundation for
theoretical analysis which would be indispensible for the studies for photo-
and electroproduction of eta mesons\cite{dytman1,ritchie} with such
spectrometers.

\section{ACKNOWLEDGMENTS}

We are particularly grateful to R. M. Davidson
and L. Zhang for many helpful discussions.
We thank many experimental colleagues around the
world for their strong interests and inputs, especially, S. Dytman,
B. Gothe, B. Krusche, B. Ritchie, B. Schoch, P. Stoler and
H. Stroeher. The research at Rensselaer has been supported by the
U. S. Department of Energy, while the research at SAL has been supported
by the Natural Sciences and Engineering Research Council of Canada. One of
us (NCM) thanks D. M. Skopik  and  E. Tomusiak for their warm hospitality
at SAL.



\newpage

\begin{appendix}

\centerline{{\bf APPENDICES}}

\section{Relation between the CGLN amplitudes and the helicity amplitudes}
In this section, for completeness, we discuss both photo- and electroproduction
amplitudes.

In the c.m. system, we quantize the initial and final spins along the
directions of $\hat{k}$ and $\hat{q}$. We choose the z-axis along
the photon momentum :
\begin{equation}
\hat{k}=\frac{\vec{k}}{|\vec{k}|}=(0,0,1)\;\;\;
\hat{q}=\frac{\vec{q}}{|\vec{q}|}=(\cos\phi\sin\theta,\sin\phi\sin\theta,
\cos\theta)
\end{equation}
The spinors of the initial and final nucleon are
\begin{eqnarray}
\chi_i^{\uparrow}= \left (
\begin{array}{c} 1 \\ 0 \end{array} \right ),\;\; & &
\chi_i^{\downarrow}=\left (
\begin{array}{c} 0\\ 1 \end{array}\right ) , \\
\chi_f^{\uparrow}= \left (
\begin{array}{c} \cos\frac{\theta}{2}  \\
\sin\frac{\theta}{2}e^{i\phi} \end{array}\right )
,\;\;&&\chi_f^{\downarrow}=\left (
\begin{array}{c} -\sin\frac{\theta}{2}e^{-i\phi} \\
\cos\frac{\theta}{2}  \end{array}\right ),
\end{eqnarray}
with
\begin{eqnarray}
\vec{\sigma}\cdot \hat{k} \chi_i^{\uparrow\;\;\downarrow}&=&
\pm \chi_i^{\uparrow\;\;\downarrow}, \\
\vec{\sigma}\cdot \hat{q} \chi_f^{\uparrow\;\;\downarrow}&=&
\pm \chi_f^{\uparrow\;\;\downarrow}.
\end{eqnarray}
Spin up would correspond, in the c.m. frame,
 to a negative helicity
and vice versa.
Explicitly, for the initial and final nucleon we have
\begin{equation}
\chi_{i,f}^{\uparrow\;\;\downarrow}=|\varrho_{i,f}=\mp 1/2>.
\end{equation}
In the case of the virtual photon, we have,
$k\cdot\varepsilon=0$;  the photon
polarization has  three independent components. To be consistent with the
photoproduction process\cite{walker}, it is convenient to take two of them to
be
\begin{equation}
\varepsilon^\mu(\lambda_\gamma)=\frac{1}{\sqrt{2}}
(0,-\lambda_\gamma,-i,0) \;\;\;\;\; \lambda_\gamma=\pm 1.
\end{equation}
The third vector is chosen,
with the normalization $\varepsilon\cdot\varepsilon=1$, to be
\begin{equation}
\varepsilon^\mu(0)=\frac{1}{\sqrt{-k^2}}(|\vec{k}|,0,0,k_0).
\end{equation}
Using the relations above and defining\cite{walker} $A=-i{\cal F}$,
Eq.(26) becomes
\begin{eqnarray}
A_{\varrho\lambda}&=&
-\frac{1}{\sqrt{2}}(\lambda_\gamma + 2 \varrho_i)
[\cos\frac{\theta}{2} \delta_{\varrho_f,-\varrho_i}
-2\varrho_f \sin\frac{\theta}{2}
e^{\textstyle 2i\varrho_f\phi}\delta_{-\varrho_f,-\varrho_i}]
({\cal F}_1 + 4 \varrho_f\varrho_i{\cal F}_2) \nonumber \\[1em]
&&+\frac{1}{\sqrt{2}}\lambda_\gamma e^{\textstyle i\lambda_\gamma\phi}
\sin\theta [\cos\frac{\theta}{2} \delta_{\varrho_f,\varrho_i}
-2\varrho_f \sin\frac{\theta}{2}
e^{\textstyle 2i\varrho_f\phi}\delta_{-\varrho_f,\varrho_i}]
(2\varrho_i {\cal F}_3 + 2\varrho_f{\cal F}_4),\nonumber \\[1em]
&&
\end{eqnarray}
for $\lambda_\gamma=\pm 1$, and
\begin{equation}
A_{\varrho\lambda}=
\sqrt{\frac{\textstyle -k^2}{\textstyle |\vec{k}|}}
[\cos\frac{\textstyle \theta}{\textstyle 2} \delta_{\varrho_f,\varrho_i}
-2\varrho_f \sin\frac{\textstyle \theta}{\textstyle 2}
e^{\textstyle 2i\varrho_f\phi}\delta_{-\varrho_f,\varrho_i}]
(2\varrho_f {\cal F}_5 + 2\varrho_i {\cal F}_6),
\end{equation}
for $\lambda_\gamma=0$, where $\lambda=\lambda_\gamma-\varrho_i$ and
$\varrho=-\varrho_f$.
By separating the $\phi$ phase factor, the following helicity
amplitudes are  defined\cite{lyth} in the usual manner:
\begin{equation}
\begin{array}{rcl}
H_1&=&e^{\textstyle -i\phi}A_{1/2\;3/2}=
-\frac{1}{\sqrt{\textstyle 2}}\sin\theta\cos
\frac{\textstyle \theta}{\textstyle 2}
({\cal F}_3 + {\cal F}_4), \\[1em]
H_2&=&A_{1/2\;1/2}=
\sqrt{\textstyle 2}\cos\frac{\textstyle \theta}{\textstyle 2}\{(
{\cal F}_2-{\cal F}_1)+\sin^2\frac{\textstyle \theta}{\textstyle 2}
({\cal F}_3-{\cal F}_4)\}, \\[1em]
H_3&=&e^{\textstyle -2i\phi}A_{-1/2\;3/2}=
\frac{1}{\sqrt{\textstyle 2}}\sin\theta
\sin\frac{\textstyle \theta}{\textstyle 2}
({\cal F}_3 - {\cal F}_4), \\[1em]
H_4&=&e^{\textstyle -i\phi}A_{-1/2\;1/2}=
\sqrt{2}\sin\frac{\textstyle \theta}{\textstyle 2}\{(
{\cal F}_2+{\cal F}_1)+
\cos^2\frac{\textstyle \theta}{\textstyle 2}({\cal F}_3+{\cal F}_4)\},\\[1em]
H_5&=&A_{-1/2\;-1/2}=
\sqrt{\frac{\textstyle -k^2}{\textstyle |\vec{k}|}}
\cos\frac{\textstyle \theta}{\textstyle 2}({\cal F}_5+{\cal F}_6),
\\[1em]
H_6&=&e^{\textstyle i\phi}A_{1/2\;-1/2}=
\sqrt{\frac{\textstyle -k^2}{\textstyle |\vec{k}|}}\sin
\frac{\textstyle \theta}{\textstyle 2}
({\cal F}_6-{\cal F}_5).
\end{array}
\end{equation}

For photoproduction, hereafter, we shall take $\lambda_\gamma=\pm 1$,
and the amplitudes $H_5$ and $H_6$ to be zero.

\section{Multipoles and helicity elements}
\subsection{Helicity Element}
The angular dependence of the ${\cal F}_i$ amplitudes, defined in
Eq.(25), can be now made explicit through their expansion, in terms of
the multipoles and the derivatives of the Legendre
polynomials $P_{\ell}(x)$ of the first kind\cite{berends,zagury,lyth}:
\begin{equation}
\begin{array}{rcl}
{\cal F}_1&=&\sum_{\ell =0}^{\infty}
[\ell M_{\ell +}+E_{\ell +}]P'_{\ell +1} +
[(\ell +1)M_{\ell -}+E_{\ell -}]P'_{\ell -1}, \\
{\cal F}_2&=&\sum_{\ell =1}^{\infty}
[(\ell +1)M_{\ell +}+\ell M_{\ell -}]P'_\ell, \\
{\cal F}_3&=&\sum_{\ell =1}^{\infty}
[E_{\ell +}-M_{\ell +}]P''_{\ell +1} +
[E_{\ell -}+M_{\ell -}+]P''_{\ell -1},  \\
{\cal F}_4&=&\sum_{\ell =2}^{\infty}
[M_{\ell +}-E_{\ell +}-M_{\ell -}-E_{\ell -}]P''_\ell .
\end{array}
\end{equation}
We note here the extra amplitudes for electroproduction, ${\cal F}_{5}$
and ${\cal F}_{6}$, given by
\begin{equation}
\begin{array}{rcl}
{\cal F}_5&=&\sum_{\ell =1}^{\infty}
[\ell S_{\ell -}-(\ell +1) S_{\ell +}]P'_\ell,  \\
{\cal F}_6&=&\sum_{\ell =0}^{\infty}
[(\ell +1)S_{\ell +}P'_{\ell +1}-\ell S_{\ell -}P'_{\ell -1}]

\end{array}
\end{equation}
The inverse relations between the multipoles and the c.m. amplitudes
${\cal F}_i$ involve projections by angular integration  are given
by Eq.(B9).

Following Jacob and Wick\cite{jacob}, the angular momentum decomposition of
the helicity amplitudes $A_{\mu\lambda}(\theta,\phi)$ is written
as\cite{amaldi}
\begin{equation}
A_{\varrho\lambda}(\theta,\phi)=\sum_{J} A_{\varrho\lambda}^{J}
(2J+1)d_{\lambda\varrho}^{J}(\theta) e^{i(\lambda-\varrho)\phi},
\end{equation}
where $\theta,\phi$ represent the angular direction of the outgoing
meson, $\varrho=-\varrho_f$ and $\lambda=\lambda_\gamma-\varrho_i$, with
$\varrho_f,\varrho_i$ being the final and initial nucleon helicities and
$\lambda_\gamma$ the photon helicity.
 For transverse photons,
$\lambda_\gamma=\pm 1$ leads to four possibilities for the initial
$\gamma N$ state
helicity $\lambda=\pm \frac{1}{2},\pm \frac{3}{2}$.
 For scalar photons,
$\lambda_\gamma=0$ and $\lambda=\pm \frac{1}{2}$.
 In total, there are
eight  helicity amplitudes $A_{\varrho\lambda}$, but the parity
symmetry\cite{jacob}
\begin{equation}
A_{-\varrho,-\lambda}(\theta,\phi)=-\exp^{i(\lambda-\varrho)(\pi-2\phi)}
A_{\varrho,\lambda}(\theta,\phi),
\end{equation}
reduces this number to four. Since the functions
$\sqrt{(2J+1)}d_{\lambda\varrho}^{J}(\theta)\exp^{i(\lambda-\varrho)\phi}$,
for different values of $J$, are mutually orthogonal and
normalized to $4\pi$,  when integrated over
$d\Omega$, the partial wave helicity amplitudes $A_{\varrho,\lambda}^{J}$
can be readily deduced from Eq.(B3):
\begin{equation}
A_{\varrho\lambda}^{J}=\frac{1}{4\pi}\int d\Omega
A_{\varrho\lambda}(\theta,\phi)d_{\lambda\varrho}^{J}
\exp^{-i(\lambda-\varrho)\phi}.
\end{equation}
Here $A_{\varrho\lambda}^{J}$ depends only on the energy and can be
combined into four independent partial wave amplitudes, often called
helicity elements
(proportional to $A_{\varrho,\lambda}^{J} \pm
A_{-\varrho,\lambda}^{J}$),
of good parity and total angular momentum $J$. These can be defined
as:
\begin{equation}
\begin{array}{rcl}
A_{\ell +}&=&-\frac{1}{\sqrt{2}}
(A_{1/2,1/2}^{J}+A_{-1/2,1/2}^{J}),  \\
A_{(\ell +1)-}&=&\frac{1}{\sqrt{2}}
(A_{1/2,1/2}^{J}-A_{-1/2,1/2}^{J}), \\
B_{\ell +}&=&\sqrt{\frac{2}{\ell (\ell +2)}}
(A_{1/2,3/2}^{J}+A_{-1/2,3/2}^{J}),  \\
B_{(\ell +1)-}&=&-\sqrt{\frac{2}{\ell (\ell +2)}}
(A_{1/2,3/2}^{J}-A_{-1/2,3/2}^{J}),\\
C_{\ell +}&=&-\frac{1}{\sqrt{2}}(A^J_{-1/2,-1/2}+A^J_{1/2,-1/2}),\\
C_{(\ell +1)-}& =&\frac{1}{\sqrt{2}}(A^{J}_{-1/2, -1/2}-A^J_{1/2, -1/2})
\end{array}
\end{equation}
where $\ell \pm$ refer to the two states with $\eta$ orbital angular
momentum $\ell$ and total angular momentum $J=\ell \pm \frac{1}{2}$. The
 four helicity elements correspond to transverse photons with
helicity $\lambda_\gamma=+1$. The $\lambda_\gamma=-1$ helicity
elements are simply related to the $\lambda_\gamma=+1$ via Eq.(B4).
The last two helicity elements refer to scalar photons
with helicity $\lambda_\gamma=0$.
 Eq. (B3) can be rewritten in
terms of the $A's,\;B's$  and the derivatives of the
Legendre polynomials of the first kind:
\begin{equation}
\begin{array}{rcl}
A_{1/2,1/2}(\theta,\phi)&=&\sqrt{2}\cos\frac{\theta}{2}
\sum_{\ell =0}^{\infty} (A_{(\ell +1)-}-A_{\ell +})
(P'_{(\ell +1)}-P'_{\ell}),  \\[0.5em]
A_{-1/2,1/2}(\theta,\phi)&=&\sqrt{2} e^{i\phi}\sin\frac{\theta}{2}
\sum_{\ell =0}^{\infty} (A_{(\ell +1)-}+A_{\ell +})
(P'_{(\ell +1)}+P'_{\ell}),  \\[0.5em]
A_{1/2,3/2}(\theta,\phi)&=&\frac{1}{\sqrt{2}} e^{i\phi}
\sin\theta\cos\frac{\theta}{2}
\sum_{\ell =1}^{\infty} (B_{(\ell +1)-}-B_{\ell +})
(P''_{(\ell +1)}-P''_{\ell}),  \\[0.5em]
A_{-1/2,3/2}(\theta,\phi)&=&\frac{1}{\sqrt{2}} e^{2i\phi}
\sin\theta\sin\frac{\theta}{2}
\sum_{\ell =1}^{\infty} (B_{(\ell +1)-}+B_{\ell +})
(P''_{(\ell +1)}+P''_{\ell}), \\
A_{-1/2, -1/2}(\theta ,\phi )& =& \sqrt{2}\cos\frac{\theta}{2}\sum^{\infty}_{
\ell =0}(C_{(\ell +1)-}-C_{\ell +})(P^{\prime}_{(\ell +1)}-P^{\prime}_{\ell}),
\\
A_{-1/2, +1/2}(\theta ,\phi )& = & -\sqrt{2}e^{-i\phi}\sin\frac{\theta}{2}
\sum^{\infty}_{\ell =0}(C_{(\ell +1)-}+C_{\ell +})(P^{\prime}_{(\ell +1)}+
P^{\prime}_{\ell}).
\end{array}
\end{equation}

\subsection{Relations between the multipoles and the helicity elements}

The relations between the multipoles and the helicity elements can now be
established by substituting Eq.(B1,B2) into Eq.(A11) and
comparing the result with relations given by Eq.(B7). Explicitly they
are given by
\begin{equation}
\begin{array}{rcl}
A_{\ell +}&=&\frac{1}{2}[\ell M_{\ell +} + (\ell+2) E_{\ell +}], \\
B_{\ell +}&=&E_{\ell +}-M_{\ell +}, \\
A_{(\ell +1)-}&=&\frac{1}{2}[(\ell +2)M_{(\ell +1)-} - \ell E_{(\ell
+1)-}], \\
B_{(\ell +1)-}&=& E_{(\ell +1)-}+M_{(\ell +1)-},\\
C_{\ell +}&=&-\sqrt{\frac{-k^2}{2|\vec{k}|}}(\ell +1) S_{\ell +}\\
C_{(\ell +1)-}&=&\sqrt{\frac{-k^2}{2|\vec{k}|}}(\ell +1) S_{(\ell +1)-}.
\end{array}
\end{equation}
Eq. (B1,B2) can be inverted to give the multipoles\cite{berends}
\begin{equation}
\begin{array}{rcl}
&\hspace*{-12.0pt}&E_{\ell +}=\frac{1}{2(\ell +1)} \int_{-1}^{+1} dx
\left [P_{\ell}{\cal F}_1 - P_{\ell +1}{\cal F}_2
+\frac{\ell}{2\ell +1}(P_{\ell -1}-P_{\ell +1}){\cal F}_3
+\frac{\ell +1}{2\ell +3}(P_{\ell}-P_{\ell +2}){\cal F}_4 \right ],
\\[1em]
&\hspace*{-12.0pt}&E_{\ell -}=\frac{1}{2\ell} \int_{-1}^{+1} dx
\left [P_{\ell}{\cal F}_1 - P_{\ell -1}{\cal F}_2
+\frac{\ell +1}{2\ell +1}(P_{\ell +1}-P_{\ell -1}){\cal F}_3
+\frac{\ell}{2\ell -1}(P_{\ell}-P_{\ell -2}){\cal F}_4 \right ], \\[1em]
&\hspace*{-12.0pt}&M_{\ell +}=\frac{1}{2(\ell +1)} \int_{-1}^{+1} dx
\left [P_{\ell}{\cal F}_1 - P_{\ell +1}{\cal F}_2
-\frac{1}{2\ell +1}(P_{\ell -1}-P_{\ell +1}){\cal F}_3 \right ], \\[1em]
&\hspace*{-12.0pt}&M_{\ell -}=\frac{1}{2\ell} \int_{-1}^{+1} dx
\left [-P_{\ell}{\cal F}_1 + P_{\ell -1}{\cal F}_2
+\frac{1}{2\ell +1}(P_{\ell -1}-P_{\ell +1}){\cal F}_3 \right ],\\
&\hspace*{-12.0pt}&S_{\ell +}=\frac{1}{2(\ell +1)} \int_{-1}^{+1} dx
\left [P_{\ell}{\cal F}_6 + P_{\ell +1}{\cal F}_5 \right ], \\[1em]
&\hspace*{-12.0pt}&S_{\ell -}=\frac{1}{2\ell} \int_{-1}^{+1} dx
\left [P_{\ell}{\cal F}_6 + P_{\ell -1}{\cal F}_5 \right ].
\end{array}
\end{equation}

\section{Observables for the $\eta$ photoproduction}

The $\eta$ photoproduction observables can be easily obtained in terms
of the helicity amplitudes defined in Eq.(A11). They are given by the
following standard expressions\cite{walker}:

(i)Differential cross section:
\begin{equation}
\frac{d\sigma}{d\Omega}=\frac{|\vec{q}|}{2|\vec{k}|}
\sum_{i=1}^{i=4}|H_i|^2.
\end{equation}

(ii) Polarized photon asymmetry:
\begin{equation}
\frac{d\sigma}{d\Omega}\Sigma=\frac{|\vec{q}|}{|\vec{k}|}
\Re e\{H_1 H_4^{\textstyle \star}-H_2 H_3^{\textstyle \star}\}.
\end{equation}

(iii) Recoil nucleon polarization in the direction
$\vec{k}\times \vec{q}$ :
\begin{equation}
\frac{d\sigma}{d\Omega}{\cal P}=-\frac{|\vec{q}|}{|\vec{k}|}
\Im m\{H_1 H_3^{\textstyle \star}+H_2 H_4^{\textstyle \star}\}.
\end{equation}

(iv) Polarized target asymmetry:
\begin{equation}
\frac{d\sigma}{d\Omega}{\cal T}=\frac{|\vec{q}|}{|\vec{k}|}
\Im m \{H_1 H_2^{\textstyle \star}+H_3 H_4^{\textstyle \star}\}.
\end{equation}

\section{Spin-$3/2$ fields}

\subsection{Spin-$3/2$ propagators}

Tt is useful to discuss some of the important
theoretical issues associated with the treatment of the spin-$3/2$ baryons.
First, the free massive spin-$3/2$ field is well known to be
consistently described by the Lagrangian\cite{moldauer,fronsdal}
\begin{equation}
{\cal L}_{Free}=\bar{\Psi}^\mu \Lambda_{\mu\nu} \Psi^\nu,
\end{equation}
with
\begin{eqnarray}
\Lambda_{\mu\nu}&=&-[(-i\partial_\lambda \gamma^\lambda +M_R)g_{\mu\nu}
-iA(\gamma_\mu\partial_\nu +\gamma_\nu \partial_\mu) \nonumber \\
&&-\frac{i}{2}(3A^2+2A+1)\gamma_\mu\partial^\lambda
\gamma_\lambda\gamma_\nu - M_R (3A^2+3A+1)],
\end{eqnarray}
where $M_R$ is the mass of the spin-$3/2$ baryon and $A$ is an arbitrary
parameter subject to the restriction $A\neq -1/2$. Physical properties
such as energy-momentum tensor \cite{nath1}
are independent of the parameter $A$,
chosen to be real here. This is due to the fact
that the free Lagrangian Eq.(D1) is invariant under the "point"
transformation\cite{johnson}
\begin{eqnarray}
\Psi^\mu &\rightarrow& \Psi^\mu+a\gamma^\mu\gamma^\nu\Psi_\nu, \\
A&\rightarrow& \frac{A-2a}{1+4a},
\end{eqnarray}
where $a\neq-1/4$, but otherwise arbitrary. Using the Euler-Lagrange
equations, the local wave equation for the spin-$3/2$ field\cite{note4}
(see also \cite{van,bdm}) can be derived\cite{rarita,aurillia}:
\begin{equation}
(i\partial_\mu \gamma^\mu -M_R) \Psi^\nu=0,
\end{equation}
with the following subsidiary conditions
\begin{eqnarray}
\gamma_\mu \Psi^\mu&=&0, \\
\partial_\mu \Psi^\mu&=&0.
\end{eqnarray}
$\Psi^\mu$ is a sixteen-component Rarita-Schwinger
vector spinor. It has a Lorentz vector index $\mu$ with a suppressed
spinor index which runs from one to four;
thus Eqs.(D6), (D7) imply a sum over the spinor index
$\beta$
\begin{equation}
(\gamma_\mu)_{\alpha\beta}\Psi^\mu_\beta=0.
\end{equation}
There are eight constraints coming from Eqs.(D6), (D7)  reducing the number
of independent components of $\Psi^\mu_\beta$ to eight (four spin
projections for the particle and the other four for the anti-particle).

The propagator for the massive spin-$3/2$ baryon can be deduced from the
equation of motion
\begin{equation}
\Lambda_{\mu\nu} \Psi^\nu=0.
\end{equation}
It satisfies the following equation
\begin{equation}
\Lambda_{\mu\lambda}G^\lambda_\nu(x,y)=\delta^4(x-y)g_{\mu\nu},
\end{equation}
where $g_{\mu\nu}$ is the metric tensor. In momentum space,
\begin{eqnarray}
&&G^\lambda_\nu(x,y)=\int \frac{d^4 p}{(2\pi)^4} G^\lambda_\nu (p)
e^{\textstyle -i(x-y)\cdot p}, \nonumber \\
&& \\
&&\Lambda_{\mu\lambda}(p)G^\lambda_\nu (p)=g_{\mu\nu}. \nonumber
\end{eqnarray}
Solving for G,
\begin{eqnarray}
G_{\mu\nu}(p)&=&\frac{\gamma\cdot p+ M_R}{p^2-M_R^2}
\left [ g_{\mu\nu} -\frac{1}{3}\gamma_\mu\gamma_\nu
-\frac{1}{3M_R}(\gamma_\mu p_\nu - \gamma_\nu p_\mu)
-\frac{2}{3M_R^2}p_\mu p_\nu \right ] \nonumber \\
&&+\frac{1}{3 M_R^2}\frac{A+1}{2A+1}
\left [ \gamma_\mu p_\nu +\frac{A}{2A+1} \gamma_\nu p_\mu
+ \left \{ \frac{1}{2}\frac{A+1}{2A+1}\gamma\cdot p
-\frac{AM_R}{2A+1} \right \}  \gamma_\mu\gamma_\nu \right ] .
\nonumber \\
&&
\end{eqnarray}
The physical properties of the free field are independent of the
parameter $A$, and we take
$A=-1$. This yields the expression for the spin-$3/2$ propagator
\begin{equation}
P_{\mu\nu}=\frac{\gamma\cdot p+ M_R}{p^2-M_R^2}
\left [ g_{\mu\nu} -\frac{1}{3}\gamma_\mu\gamma_\nu
-\frac{1}{3M_R}(\gamma_\mu p_\nu - \gamma_\nu p_\mu)
-\frac{2}{3M_R^2}p_\mu p_\nu \right ].
\end{equation}

We refer the reader to a discussion\cite{bdm} of  erroneous choices
of the spin-$3/2$ propagators  adopted by some recent
works.

\subsection{Spin projection operators for spin-$3/2$ field}

The spin projection operators are given by\cite{van,bdm}

\begin{eqnarray}
(P^{3/2})_{\mu\nu}&=&g_{\mu\nu}-\frac{1}{3}\gamma_\mu \gamma_\nu-
\frac{1}{3p^2}(\gamma\cdot p\gamma_\mu p_\nu +p_\mu \gamma_\nu
\gamma\cdot p), \nonumber \\
(P_{11}^{1/2})_{\mu\nu}&=&\frac{1}{3}\gamma_\mu\gamma_\nu-\frac{p_\mu
p_\nu}{p^2}+\frac{1}{3p^2}(\gamma\cdot p\gamma_\mu p_\nu +p_\mu
\gamma_\nu\gamma\cdot p), \nonumber \\
(P_{22}^{1/2})_{\mu\nu}&=&\frac{p_\mu p_\nu}{p^2}, \\
(P_{12}^{1/2})_{\mu\nu}&=&\frac{1}{\sqrt{3} p^2}
(p_\mu p_\nu-\gamma\cdot p\gamma_\mu p_\nu), \nonumber \\
(P_{21}^{1/2})_{\mu\nu}&=&\frac{1}{\sqrt{3} p^2}
(\gamma\cdot p\gamma_\nu p_\mu-p_\mu p_\nu).\nonumber
\end{eqnarray}
These satisfy the orthonormality conditions
\begin{equation}
(P_{ij}^I)_{\mu\lambda}(P_{kl}^J)^{\lambda\nu}=\delta^{IJ}\delta_{jk}
(P_{il}^I)_\mu^\nu,
\end{equation}
and the sum rule for the projection operators
\begin{equation}
(P^{3/2})_{\mu\nu}+(P_{11}^{1/2}){\mu\nu}+(P_{22}^{1/2}){\mu\nu}=g_{\mu\nu}.
\end{equation}
The following properties are also useful
\begin{eqnarray}
\gamma\cdot p P_{ij}^{1/2}&=&\pm P_{ij}^{1/2} \gamma\cdot p, \\
\gamma\cdot p P^{3/2}&=&P^{3/2} \gamma\cdot p,
\end{eqnarray}
where $+$ for $i=j$ and $-$ for $i\neq j$.

\section{Spin-$3/2$ isospin-$1/2$ odd parity contribution to the
invariant amplitudes}

The invariant amplitudes for the $s$-channel $C_1$ and $C_2$ couplings can be
expressed in the form
\begin{equation}
A_i=A_{i,P}+A_{i,NP}.
\end{equation}

The pole $(P)$ and the non-pole $(NP)$ terms for $C_1$ coupling are
given by
\begin{eqnarray}
A_{1,P}&=&\frac{C_1}{8(s-M_R^2)}[4t+\frac{4M^2}{3M_R^2}(M_R^2-M^2+\mu^2)-
\frac{4M}{3M_R}(M_R^2-M^2+2\mu^2)],\nonumber \\
A_{2,P}&=&\frac{C_1}{8(s-M_R^2)}[-8],\nonumber \\
A_{3,P}&=&\frac{C_1}{8(s-M_R^2)}[-2M+4M_R-\frac{4M^2}{3M_R}+
\frac{2M}{3M_R^2}(M_R^2-2M^2+2\mu^2)],\nonumber \\
A_{4,P}&=&\frac{C_1}{8(s-M_R^2)}[6M-4M_R-\frac{4M^2}{3M_R}+
\frac{2M}{3M_R^2}(M_R^2-2M^2+2\mu^2)], \nonumber \\
&& \\
A_{1,NP}&=&\frac{C_1}{12 M_R^2}[2M(M-M_R)+\beta
(\alpha-1)(s-M^2)-2\mu^2\beta], \nonumber \\
A_{2,NP}&=&0, \nonumber \\
A_{3,NP}&=&\frac{C_1}{12 M_R^2}[2\alpha\beta
M_R-M(\alpha+\beta+\alpha\beta-1)], \nonumber \\
A_{4,NP}&=&\frac{C_1}{12 M_R^2}[2\alpha\beta
M_R-M(\alpha+\beta+\alpha\beta-1)], \nonumber
\end{eqnarray}
with
\begin{eqnarray}
 \alpha=1+4Z, &\beta=1+4Y,&
C_1=\frac{e(k_R^{s,(1)} \pm k_R^{v,(1)})f_{\eta NR}}{2M\mu}.
\end{eqnarray}

The pole and the non-pole terms due to $C_2$ coupling are as follows:
\begin{eqnarray}
A_{1,P}&=&\frac{C_2}{12(s-M_R^2)} [2M(3t-2\mu^2)+
\frac{1}{M_R}(s+M^2)(s-M^2+\mu^2) \nonumber \\
       && -2M_R(s-M^2)], \nonumber \\
A_{2,P}&=&\frac{C_2}{12(s-M_R^2)}[-6(M_R+M)],\nonumber \\
A_{3,P}&=&\frac{C_2}{12(s-M_R^2)}[(3t-2\mu^2)+\frac{M}{M_R}(s-M^2+\mu^2)
+5(s-M^2)], \nonumber \\
A_{4,P}&=&\frac{C_2}{12(s-M_R^2)}[(3t-2\mu^2)+\frac{M}{M_R}(s-M^2+\mu^2)
-(s-M^2)], \nonumber  \\
&& \\
A_{1,NP}&=&\frac{C_2}{24 M_R^2}
[(s-M^2)\{ M\delta(\alpha-3)
+M_R(2\delta\alpha-\alpha-1)\}-4M\mu^2\delta],
\nonumber \\
A_{2,NP}&=&0, \nonumber \\
A_{3,NP}&=&\frac{C_2}{24 M_R^2}[(s-M^2)(\alpha-1)\delta-2\mu^2\delta],
\nonumber \\
A_{4,NP}&=&\frac{C_2}{24 M_R^2}[(s-M^2)(\alpha-1)\delta-2\mu^2\delta],
\nonumber
\end{eqnarray}
with
\begin{eqnarray}
\alpha=1 + 4Z, &\delta=1 + 2X, &
C_2=-\frac{e (k_R^{s,(2)} \pm k_R^{v,(2)}) f_{\eta NR}}{4M^2\mu}.
\end{eqnarray}

The $u$-channel invariant amplitudes are obtained from the
$s$-channel ones through crossing relations as given by Eq.(22).
\end{appendix}


\newpage

Fig. 1: {Feynman diagrams for the  $\eta$ photoproduction.
(a), (b) The direct ($s$-channel) and crossed ($u$-channel) $PS$ Nucleon Born
contributions; (c) the
equivalence breaking  contribution; (d) the $t$-channel $\rho^\circ$ and
$\omega$ vector meson exchanges; (e), (f) the  $s$- and $u$-channel nucleon
resonance excitations.}

Fig. 2: {Differential cross section and recoil nucleon polarization
at c.m. angle of $90^\circ\pm 9^{\circ}$  for the reaction $\gamma p
\rightarrow \eta p$ as a function of the photon lab energy, a typical example
from the
``old" data set.  The  data
set of Homma {\em et al.} \protect\cite{homma} are marked with circles. The
dots
correspond to the older data\protect\cite{genzel}. The Fit A, shown by the
solid
lines, uses the PDG-92 resonance parameters. The
dashed curve excludes the $N^{*}(1535)$ contribution.  Predicted
polarized target and photon asymmetries, for which there are no data,
are also shown.  }

Fig. 3: {Calculated observables using the $KOCH$\protect\cite{koch}
resonance parameters. See Fig. 2
for explanations.}

Fig. 4: {Calculated observables using the $BAKR$\protect\cite{baker}
resonance parameters. See Fig. 2
for explanation.}

Fig. 5: {Predicted total cross section from the Fit A
as a function of incident photon lab energy, for
the process $\gamma p \rightarrow \eta p$. The curves
correspond to the different resonance parameters: solid: $PDG92$, dash:
$KOCH$, dotted: $BAKR$ and data are from Ref. \protect\cite{baker}.}

Fig. 6: {Predicted angular distributions
for the  photon lab energy $E_\gamma$ = $729\,MeV$ and $753\,MeV$.
The curves correspond to the prediction of the Fit A. The circles and diamonds
are the new Bates 1993 data\protect\cite{dytman}. }

Fig. 7: { Predicted angular distributions from our Fit C  for
$E_{\gamma}$ = $729\,MeV$ and
$753\,MeV$, compared with the Bates data \protect\cite{dytman}.}

Fig. 8:{ Angular distributions, predicted from our Fit A,
for the  photon lab energy $E_\gamma=722.5$, $737.5$, $752.5$, $767.5$ and
$782.5\,MeV$ [solid curves].
The data  are preliminary, from the Mainz experiment\protect\cite{krusche},
with arbitrary normalization; these have been
normalized to our prediction by a factor of 0.75 with respect to the nominal
values in Ref. \cite{krusche}.}

Fig. 9:  {The  predicted total cross section of our  Fit A
compared with the preliminary  Mainz data\protect\cite{krusche},
with arbitrary normalization; these have been normalized by
a factor 0.75, as in Fig. 8.  }

\newpage

\begin{table}
\caption{Coupling constants of vector mesons considered in this work
\protect\cite{hoehler,grein,dolinsky}. $\Gamma$ is  the full width.
The mass of the vector meson is shown in parentheses.}
\label{table1}
\begin{center}
\begin{tabular}{lcccc}
&&&& \\
Vector meson &  $\Gamma (MeV)$ &
$\Gamma_{V\rightarrow \eta\gamma}(keV)$ & $g_v$ & $g_t$ \\[1em] \hline
&&&& \\
$\rho(770 MeV)$           & $153 \pm 2$ & $62 \pm 17$
& $2.63 \pm 0.38$& $16.05 \pm 0.82$ \\[1em]
$\omega(782 MeV)$          & $8.5 \pm 0.1$ & $6.1 \pm 2.5$
& $10.09 \pm 0.93$ & $1.42 \pm 1.99$ \\[1em]
\end{tabular}
\end{center}
\end{table}

\begin{table}
\vspace{1.5in}
\caption{Summary of the properties of the baryon resonances
considered in this work. $J^\pi$ is the spin-parity, $\Gamma$ is the
total width. The numbers in parenthesis as well as the ones in the last
column correspond to the nominal values used by the Particle Data
Group\protect\cite{pdg}.
The first row correspond to PDG 1990 ($PDG90$) and the second to PDG 1992
($PDG92$). We use the notation $L_{2I\!2J}$ used
in $\pi N$ scattering: the resonance, once produced, decays into $\pi N$ with a
relative orbital angular momentum $L$, isospin $I$ and total angular
momentum $J$.}
\label{table2}
\begin{center}
\begin{tabular}{lccccc}
&&&&& \\
Resonance & $J^\pi$ & $L_{2I\!2J}$ & $Mass\,(MeV)$ & $\Gamma \,(MeV)$ &
$\Gamma_{N^{\textstyle \star} \rightarrow N\eta}\, (MeV)$ \\[1em] \hline
&&&&& \\
$N^{\textstyle \star}(1440)$
&  $1/2^+$     & $P_{11}$ & $1400-1480$  & $120-350\;(200)$ & Not given \\
&              &          & $1430-1470$  & $250-450\;(350)$ & Not given
\\[1em] \hline
$N^{\textstyle \star}(1520)$
&  $3/2^-$     & $D_{13}$ & $1510-1530$ & $100-140\;(125)$ & $\sim 0.125$ \\
&              &          & $1515-1530$ & $110-135\;(120)$ & $\sim 0.12$
\\[1em] \hline
$N^{\textstyle \star}(1535)$
&  $1/2^-$     & $S_{11}$ & $1520-1560$ & $100-250\;(150)$ & $67.5-82.5$ \\
&              &          & $1520-1555$ & $100-250\;(150)$ & $45-75$
\\[1em] \hline
$N^{\textstyle \star}(1650)$
&  $1/2^-$     & $S_{11}$ & $1620-1680$ & $100-200\;(150)$ & $\sim 2.25$ \\
&              &          & $1640-1680$ & $145-190\;(150)$ & $\sim 1.5$
\\[1em] \hline
$N^{\textstyle \star}(1710)$
&  $1/2^+$     & $P_{11}$ & $1680-1740$ & $90-130\;(110)$ & $\sim 27.5$ \\
&              &          & $1680-1740$ & $50-250\;(100)$ & $\sim 20-40$
\end{tabular}
\end{center}
\vspace{1.5in}
\end{table}

\begin{table}
\caption{The strong decay widths and the helicity amplitudes as given
by the $PDG92$,
compared with the
constituent quark model calculations(QM). For each $N^{*}$ the first row
corresponds to  the  calculation of Koniuk and
Isgur\protect\cite{koniuk}, the second row  to  the
calculation of Capstick and Roberts\protect\cite{capstick}.}
\label{table3}
\begin{center}
\begin{tabular}{lcccccc}
&&&&&& \\
$$
&\multicolumn{2}{c}{$\sqrt{\Gamma_\eta}\;(MeV^{1/2})$}
&\multicolumn{2}{c}{$A_{1/2}^p\;(10^{-3} GeV^{1/2})$}
&\multicolumn{2}{c}{$A_{3/2}^p\;(10^{-3} GeV^{1/2})$} \\[1em]
& $PDG$ & QM & $PDG$ & QM & $PDG$ & QM \\[1em]
 \hline
&&&&&& \\
$N^{*}(1440)$ & $--$ & $+small$ & $-68\pm 5$ & $-24$
& $--$ & $--$ \\[1em]
& $--$ & $0.0^{+1.0}_{-0.0}$ &  & 4 &$--$ &$--$ \\[1em] \hline
$N^{*}(1520)$ & $+0.35$ & $+0.4$ & $-23\pm 9$ & $-23$
& $+163\pm 8$ & $+128$     \\[1em]
              & & $+0.4^{+2.9}_{-0.4}$ & & $-15$ && $+134$ \\[1em] \hline
$N^{*}(1535)$ & $6.7-8.7$ & $+5.2$ & $+74\pm 11$ & $+147$
& $--$ & $--$ \\[1em]
& & $14.6^{+0.7}_{-1.3}$ & & 76 &$--$ &$--$ \\[1em]\hline
$N^{*}(1650)$ & $-1.22$ & $-1.5$ & $+48\pm 16$ & $+88$
& $--$ & $--$ \\[1em]
& & $-7.8^{+0.1}_{-0.0}$ & & 54&$--$ &$--$ \\[1em]\hline
$N^{*}(1710)$ & $4.47-6.32 $ & $+2.9$ & $+5\pm 16$ & $-47$
& $--$ & $--$  \\[1em]
& & $5.7\pm 0.3$ & & $+13$ &$--$ &$--$ \\[1em]
\end{tabular}
\end{center}
\end{table}

\begin{table}
\vspace{1.5in}
\caption{The masses and widths (in $MeV$)
of the baryon resonances considered in this work.
$KOCH$ and $CUTK$ resonance parameters are determined from
$\pi N\rightarrow \pi N$ scattering in the analysis by
Koch\protect\cite{koch}
and Cutkosky\protect\cite{cutkosky} respectively. $BAKR$ refers to Baker
analysis\protect\cite{baker} of
the data on the reaction $\pi^- p\rightarrow \eta n$. PDG92 are the nominal
values given by  Particle Data Group (1992).}
\label{table4}
\begin{center}
\begin{tabular}{lcccc}
&&&& \\
$$   &    $KOCH$  & $CUTK$  &  $BAKR$& $PDG92$ \\[1em] \hline
&&& \\
$N^{*}(1440)$ &    $1410,135$   & $1440,340$
& $1472,113$  & 1440,350 \\[1em]
$N^{*}(1520)$ &    $1519,114$   & $1525,120$
& $1520,183$  & 1520,120  \\[1em]
$N^{*}(1535)$ &    $1526,120$   & $1550,240$
& $1517,180$  & 1535,150  \\[1em]
$N^{*}(1650)$ &    $1670,180$   & $1650,150$
& $1670,200$  & 1650,150  \\[1em]
$N^{*}(1710)$ &    $1723,120$   & $1700,90$
& $1690,97$   & 1710,100 \\[1em]
\end{tabular}
\end{center}
\vspace{1.5in}
\end{table}

\begin{table}
\caption{Couplings obtained by fitting the old data base for eta
photoproduction,
using  different  sets of masses and widths of
the resonances. The cutoff used for the vector meson form factor (Eq.(60))
is indicated by $\Lambda^2$. The off-shell parameter at the strong vertex is
fixed to $\alpha=-1$. There are 137 differential cross section and 7
polarization data points below $E_\gamma=1.2\;GeV$ in the old base.
Eight parameters are fitted. The $\eta NN$ and $\eta NN^{*}$ couplings are
pseudoscalar. The corresponding helicity amplitudes are listed in units
of $10^{-3} GeV^{1/2}$.
The quantity $\xi$ characteristic of the photoexcitation of the dominant
resonance $N^{*}(1535)$ and its decay into the $\eta N$ channel is also
given in units $10^{-4} MeV^{-1}$.}
\label{table5}
\small
\begin{center}
\begin{tabular}{lcccccc}
&&&&&& \\
& \multicolumn{2}{c}{$PDG92$}
& \multicolumn{2}{c}{$KOCH$}
& \multicolumn{2}{c}{$BAKR$}
\\
$\Lambda^2(GeV^2)$ & $2.0$ & $1.2$ & $2.0$ & $1.2$ & $2.0$ & $1.2$
\\[1em] \hline

$g_\eta$
& $2.6$  & $4.1$ & $6.8$ & $5.8$ & $4.2$ & $4.1$  \\[1em]

$[\sqrt{\Gamma_\eta} A_{1/2}]_{S1}$
& $26.6$ & $26.2$ & 20.0& 20.3& 26.7& 26.3 \\
$[A_{1/2}]_{S1}$
& $97.0$       & $95.6$ & 81.7& 82.9& 104.9& 103.3
\\[1em]

$[\sqrt{\Gamma_\eta} A_{1/2}]_{S2}$
& $-6.9$ & $-7.0$ & $-3.8$ & $-4.1$ & $-17.3$ & $-14.3$ \\
$[A_{1/2}]_{S2}$
& $179.4$     & $180.9$ & 72.4& 78.5& 315.4& 260.8\\[1em]

$[\sqrt{\Gamma_\eta} A_{1/2}]_{P1}$
& $-3.8$ & $-0.08$ & $0.03$ & $0.03$ & $-0.1$ & $-0.2$ \\
$[A_{1/2}]_{P1}$
& $-17.4$       & $-0.4$ & $0.1$ & $0.1$ & $-1.6$ & $-2.6$ \\[1em]

$[\sqrt{\Gamma_\eta} A_{1/2}]_{P2}$
& $3.9$  & $2.9$ & $4.1$ & $5.0$ & $4.3$ & $3.4$ \\
$[A_{1/2}]_{P2}$
& $22.7$      & $16.7$ & 23.7& 28.9& 27.6& 22.0\\[1em]

$[\sqrt{\Gamma_\eta} A_{1/2}]_{D1}$
& $-0.1$ & $-0.1$ & $-0.2$ & $-0.2$ & $-0.07$ & $-0.09$ \\
$[A_{1/2}]_{D1}$
& $-9.5$      & $-9.3$ & $-15.2$ & $-16.9$ & $-5.3$ & $-6.5$\\[1em]

$[\sqrt{\Gamma_\eta} A_{3/2}]_{D1}$
& $0.8$  & $0.8$ & $1.2$ & $1.4$ & $0.5$ & $0.7$ \\
$[A_{3/2}]_{D1}$
& $72.4$     & $70.4$ & 116.1& 128.9& 40.3& 49.3\\[1em]

$\beta$
& $1.4$  & $0.9$ & $1.0$ & $1.0$ & $1.1$ & $0.8$ \\
$\delta$
& $1.4$  & $1.0$ & $1.4$ & $1.3$ & $1.4$ & $1.3$ \\[1em]

$\chi^2(XS)$
& $239.8$  & $226.9$ & 295.0& 266.5& 259.8& 242.1 \\
$\chi^2(POL)$
& $30.5$   & $12.2$ & $12.1$ & $13.8$ & $29.6$ & $16.6$ \\
$\chi^2_{TOT}/DF$
& $2.0$   & $1.8$ & $2.3$ & $2.1$ & $2.1$ & $1.9$ \\[1em] \hline\hline

$\xi$
& $2.22\pm 0.15$ & $2.19\pm 0.15$ & $2.19 \pm 0.18$
& $2.23\pm 0.17$ & $2.07\pm 0.14$ & $2.04\pm 0.14$

\end{tabular}
\end{center}
\end{table}

\begin{table}
\caption{Same as Table \protect\ref{table5}, with $\alpha=1$.}
\begin{center}
\label{table6}
\small
\begin{tabular}{lcccccc}
&&&&&& \\
& \multicolumn{2}{c}{$PDG92$}
& \multicolumn{2}{c}{$KOCH$}
& \multicolumn{2}{c}{$BAKR$} \\
$\Lambda^2 (GeV^2)$& $2.0$ & $1.2$ & $2.0$ & $1.2$ & $2.0$ & $1.2$
\\[1em] \hline

$g_\eta$
& $5.8$    & $4.1$ &  $6.3$& 5.1& 6.1& 4.1  \\[1em]

$[\sqrt{\Gamma_\eta} A_{1/2}]_{S1}$
& $25.8$    & $26.1$ & $20.2$ & $20.8$ & $26.0$& 26.6  \\
$[A_{1/2}]_{S1}$
& $94.1$       & $95.5$ & $82.5$ & $84.9$ & $102.0$ & $104.5$  \\[1em]

$[\sqrt{\Gamma_\eta} A_{1/2}]_{S2}$
& $-6.1$    & $-7.0$ & $-5.5$ & $-6.7$ & $-13.3$ & $-15.1$  \\
$[A_{1/2}]_{S2}$
& $156.7$       & $180.3$ & $105.9$ & $128.9$ & $243.0$ & $275.7$  \\[1em]

$[\sqrt{\Gamma_\eta} A_{1/2}]_{P1}$
& $-1.4$    & $-1.4$ & $-2.3$ & $0.03$ & $-0.06$ & $-0.3$  \\
$[A_{1/2}]_{P1}$
& $-6.2$       & $-6.4$ & $-7.4$ & $0.1$ & $-0.7$ & $-3.6$  \\[1em]

$[\sqrt{\Gamma_\eta} A_{1/2}]_{P2}$
& $3.8$    & $3.9$ & $ 7.7$& 7.1& 3.3& 3.4 \\
$[A_{1/2}]_{P2}$
& $22.2$       & $22.6$ & 44.4&41.0&21.4&21.9  \\[1em]

$[\sqrt{\Gamma_\eta} A_{1/2}]_{D1}$
& $-0.2$    & $-0.2$ & $-0.3$ & $-0.2$ & $-0.2$ & $-0.2$  \\
$[A_{1/2}]_{D1}$
& $-17.6$       & $-14.8$ & $-24.2$ & $-20.4$ & $-14.1$ & $-11.7$  \\[1em]

$[\sqrt{\Gamma_\eta} A_{3/2}]_{D1}$
& $1.5$    & $1.2$ & $2.0$ & $1.7$ & $1.4$ & $1.2$  \\
$[A_{3/2}]_{D1}$
& $133.6$       & $112.6$ & 184.8& 155.6& 106.7& 89.2 \\[1em]

$\beta$
& $-0.3$    & $0.09$ & $-0.3$ & $-0.06$ & $-0.3$ & $0.2$  \\
$\delta$
& $-3.0$    & $-1.5$ & $-4.0$ & $-2.6$ & $-1.5$ & $0.8$  \\[1em]

$\chi^2(XS)$
& $229.2$    & $221.0$ & $254.7$ & $258.4$& 244.0& 235.7  \\
$\chi^2(POL)$
& $14.4$    & $12.5$ & $16.2$ & $17.3$ & $15.6$& 13.7  \\
$\chi^2_{TOT}/DF$
& $1.8$    & $1.7$ & $2.0$ & $2.0$ & $1.9$ & $1.8$  \\[1em] \hline\hline

$\xi$
& $2.16\pm 0.15$  & $2.19\pm 0.15$
& $2.22\pm 0.18$ & $2.28\pm 0.17$
& $2.02\pm 0.14$ & $2.07\pm 0.14$
\end{tabular}
\end{center}
\end{table}

\begin{table}
\caption{Results  of fitting the old data base
using the pseudovector $\eta NN$ coupling, with the pseudoscalar $\eta NN^{*}$
couplings.
Only the values of
$\alpha$ that give the lowest $\chi^2/DF$ are shown. In the fit, the value of
$\alpha$ is fixed. Other notations are
the same as in Table \protect\ref{table5}.}
\begin{center}
\label{table7}
\small
\begin{tabular}{lcccccc}
&&&&&& \\
& \multicolumn{2}{c}{$PDG92$}
& \multicolumn{2}{c}{$KOCH$}
& \multicolumn{2}{c}{$BAKR$} \\
$\Lambda^2 (GeV^2)$& $1.2$ & $2.0$ & $1.2$ & $2.0$ & $1.2$ & $2.0$
\\[1em] \hline

$g_\eta$
& $4.4$    & $3.4$ & $5.0$ & $6.2$ & $4.3$&4.6  \\[1em]

$[\sqrt{\Gamma_\eta} A_{1/2}]_{S1}$
& $26.1$    & $26.6$ & $20.5$ & $20.1$ & $26.5$&26.6  \\
$[A_{1/2}]_{S1}$
& $95.2$       & $97.0$ & $83.6$ & $81.9$ & $104.1$& 104.4  \\[1em]

$[\sqrt{\Gamma_\eta} A_{1/2}]_{S2}$
& $-7.0$    & $-7.2$ & $-6.5$ & $-5.6$ & $-14.9$ & $-13.9$  \\
$[A_{1/2}]_{S2}$
& $181.5$       & $185.5$ & $125.3$ & $107.0$ & 271.9& 254.2  \\[1em]

$[\sqrt{\Gamma_\eta} A_{1/2}]_{P1}$
& $-1.8$    & $-2.0$ & $-3.4$ & $-4.3$ & $-0.3$ & $-0.3$  \\
$[A_{1/2}]_{P1}$
& $-8.4$       & $-9.2$ & $-10.8$ & $-13.6$ & $-3.2$ & $-3.2$  \\[1em]

$[\sqrt{\Gamma_\eta} A_{1/2}]_{P2}$
& $3.8$    & $3.1$ & $8.0$ & $7.9$ &  $3.3$&3.0  \\
$[A_{1/2}]_{P2}$
& $21.9$       & $17.9$ & $46.4$ & $45.7$ & 21.5&19.4  \\[1em]

$[\sqrt{\Gamma_\eta} A_{1/2}]_{D1}$
& $-0.1$    & $-0.2$ & $-0.2$ & $-0.2$ & $-0.1$ & $-0.2$  \\
$[A_{1/2}]_{D1}$
& $-11.9$   & $-18.3$ & $-19.1$ & $-19.0$ & $-8.4$ & $-18.2$  \\[1em]

$[\sqrt{\Gamma_\eta} A_{3/2}]_{D1}$
& $1.0$    & $1.5$ & $1.6$ & $1.5$ & $0.9$ & $1.9$  \\
$[A_{3/2}]_{D1}$
& $90.5$   & $138.6$ & $146.1$ & $145.1$ & 64.1& 138.5  \\[1em]

$\alpha$
& $1.0$    & $0.0$ & $1.0$ & $1.0$ & $1.0$ & $0.0$  \\
$\beta$
& $0.9$    & $1.3$ & $0.5$ & $0.4$ & 1.2&1.2  \\
$\delta$
& $-10.2$    & $0.04$ & $-9.4$ & $-12.6$ & $-8.7$& 0.07  \\[1em]

$\chi^2(XS)$
& $224.2$    & $230.1$ & $255.6$ & $268.9$ & $240.7$ & $236.0$  \\
$\chi^2(POL)$
& $13.3$    & $11.3$ & $17.5$ & $19.2$ & $15.8$ & $12.0$  \\
$\chi^2_{TOT}/DF$
& $1.8$    & $1.8$ & $2.0$ & $2.1$ & $1.9$ & $1.8$  \\[1em] \hline\hline

$\xi$
& $2.18\pm 0.15$ & $2.22\pm 0.15$
& $2.25\pm 0.18$ & $2.20\pm 0.18$
& $2.06\pm 0.14$ & $2.06\pm 0.14$
\end{tabular}
\end{center}
\end{table}

\begin{table}
\caption{Fitted couplings from the old data base
 using Cutkosky ($CUTK$) resonance
parameters. The $\eta NN^{*}$ coupling is pseudoscalar, while
$\eta NN$ coupling chosen is shown.  Only lowest $\chi^2/DF$ are given.}
\begin{center}
\label{table8}
\small
\begin{tabular}{lcccc}
&&&& \\
$\Lambda^2 (GeV^2)$& $2.0$ & $1.2$ & $1.2$ & $1.2$ \\
$\eta N N$ & $PS$ & $PS$ & $PV$ & $PS$ \\[1em] \hline

$g_\eta$
& $0.06$    & 2.0& 0.9& 1.7 \\[1em]

$[\sqrt{\Gamma_\eta} A_{1/2}]_{S1}$
& $45.0$    & $44.2$ & $44.7$ & 44.2 \\
$[A_{1/2}]_{S1}$
& $176.7$ & $173.6$ & $175.7$ & $173.7$ \\[1em]

$[\sqrt{\Gamma_\eta} A_{1/2}]_{S2}$
& $-13.1$    & $-11.0$ & $-12.8$ & $-11.2$ \\
$[A_{1/2}]_{S2}$
& $277.0$       & $232.4$ & $269.0$ & $235.5$ \\[1em]

$[\sqrt{\Gamma_\eta} A_{1/2}]_{P1}$
& $-2.4$    & $-1.9$ & $-1.3$ & $-3.1$ \\
$[A_{1/2}]_{P1}$
& $-10.9$       & $-8.6$ & $-6.0$ & $-14.3$ \\[1em]

$[\sqrt{\Gamma_\eta} A_{1/2}]_{P2}$
& $1.8$    & $0.6$ & $1.7$ & $0.9$ \\
$[A_{1/2}]_{P2}$
& $12.2$       & $4.3$ & $11.3$ & $5.9$ \\[1em]

$[\sqrt{\Gamma_\eta} A_{1/2}]_{D1}$
& $-0.07$    & $-0.04$ & $-0.07$ & $-0.008$ \\
$[A_{1/2}]_{D1}$
& $-6.0$       & $-3.5$ & $-6.8$ & $-0.8$ \\[1em]

$[\sqrt{\Gamma_\eta} A_{3/2}]_{D1}$
& $0.5$    & $0.3$ & $ 0.6$ & $0.06$ \\
$[A_{3/2}]_{D1}$
& $44.5$       & $26.0$ & $50.5$ & $5.6$ \\[1em]

$\alpha$
& $2.0$    & $1.0$ & $2.0$ & $0.0$ \\
$\beta$
& $-0.04$    & $-0.09$ & $0.6$ & $0.2$ \\
$\delta$
& 5.1& 9.9& 3.8& 9.9 \\[1em]

$\chi^2(XS)$
& $217.3$    & $220.1$ & $214.9$ & $225.6$ \\
$\chi^2(POL)$
& $11.8$    & $11.6$ & $11.5$ & $11.6$ \\
$\chi^2_{TOT}/DF$
& $1.7$    & $1.7$ & $1.7$ & $1.8$ \\[1em] \hline\hline

$\xi$
& $2.21\pm 0.11$ & $2.17\pm 0.12$ & $2.20\pm 0.11$
& $2.17\pm 0.12$

\end{tabular}
\end{center}
\end{table}

\begin{table}
\caption{Fitted parameters using the old data base and the
 $PV$ coupling at the  $\eta NN$ and
$\eta N N^{\textstyle \star}$ vertices. The cutoff for the vector meson
form factor is fixed at $1.2 GeV^2$.}
\label{table9}
\begin{center}
\small
\begin{tabular}{lcccc}
&&&& \\
& $PDG92$ & $KOCH$ & $BAKR$ & $CUTK$
\\[1em] \hline

$g_\eta$
& $3.7$   & $5.3$ & $4.4$ & $2.2$ \\[1em]

$[\sqrt{\Gamma_\eta} A_{1/2}]_{S1}$
& $26.0$   & $20.1$ & $25.5$ & $43.3$ \\
$[A_{1/2}]_{S1}$
& $95.0$  & $82.2$ & $100.2$ & $169.9$ \\[1em]

$[\sqrt{\Gamma_\eta} A_{1/2}]_{S2}$
& $-9.1$   & $-6.4$ & $-14.6$ & $-11.8$ \\
$[A_{1/2}]_{S2}$
& $234.7$  & $122.7$ & $265.7$ & $248.6$ \\[1em]

$[\sqrt{\Gamma_\eta} A_{1/2}]_{P1}$
& $-0.4$   & $-2.3$ & $-0.2$ & $-2.2$ \\
$[A_{1/2}]_{P1}$
& $-1.7$  & $-7.4$ & $-2.3$ & $-10.0$ \\[1em]

$[\sqrt{\Gamma_\eta} A_{1/2}]_{P2}$
& $2.0$   & $7.7$ & $3.0$ & $0.7$ \\
$[A_{1/2}]_{P2}$
& $11.7$  & $44.4$ & $19.2$ & $4.4$ \\[1em]

$[\sqrt{\Gamma_\eta} A_{1/2}]_{D1}$
& $-0.04$   & $-0.2$ & $-0.1$ & $-0.007$ \\
$[A_{1/2}]_{D1}$
& $-3.8$  & $-19.1$ & $-8.6$ & $-0.6$ \\[1em]

$[\sqrt{\Gamma_\eta} A_{3/2}]_{D1}$
& $0.3$   & $1.6$ & $0.9$ & $0.05$ \\
$[A_{3/2}]_{D1}$
& $28.8$  & $145.9$ & $65.6$ & $4.5$ \\[1em]

$\alpha$
& $1.0$   & $1.0$ & $1.0$ & $1.0$ \\
$\beta$
& $2.5$   & $0.4$ & $0.9$ & $4.7$ \\
$\delta$
& $-12.9$   & $-8.6$ & $-6.0$ & $4.2$ \\[1em]

$\chi^2(XS)$
& $228.2$   & $247.3$ & $232.6$ & $229.0$ \\
$\chi^2(POL)$
& $14.2$   & $16.0$ & $14.2$ & $11.5$ \\
$\chi^2_{TOT}/DF$
& $1.8$   & $2.0$ & $1.8$ & $1.8$ \\[1em]\hline\hline

$\xi$
& $2.18\pm 0.14$  & $2.21\pm 0.15$ & $1.98\pm 0.12$ & $2.13\pm 0.11$

\end{tabular}
\end{center}
\end{table}

\begin{table}
\caption{Fitting the old data base together with the new Bates data
\protect\cite{dytman}.}
\label{table10}
\begin{center}
\small
\begin{tabular}{lcccc}
&&&& \\
$\Lambda^{2}$ & 2.0& 1.2& 2.0& 1.2
\\[1em] \hline

$g_\eta$
& 5.9& 4.6& 5.9& 4.3 \\[1em]

$[\sqrt{\Gamma_\eta} A_{1/2}]_{S1}$
& 26.2& 26.3& 26.2& 26.4 \\
$[A_{1/2}]_{S1}$
& 95.5& 96.0& 95.8& 96.2 \\[1em]

$[\sqrt{\Gamma_\eta} A_{1/2}]_{S2}$
& $-5.5$   & $-6.8$ & $-6.4$ & $-7.4$ \\
$[A_{1/2}]_{S2}$
& 142.9& 176.7& 164.6& 192.2 \\[1em]

$[\sqrt{\Gamma_\eta} A_{1/2}]_{P1}$
& 0.7& 1.1& 0.5& 0.2 \\
$[A_{1/2}]_{P1}$
& 3.3& 4.9& 2.5& 1.1 \\[1em]

$[\sqrt{\Gamma_\eta} A_{1/2}]_{P2}$
& 2.1& 2.8& 3.1& 3.6\\
$[A_{1/2}]_{P2}$
& 12.1& 16.4& 18.1& 20.7 \\[1em]

$[\sqrt{\Gamma_\eta} A_{1/2}]_{D1}$
& $-0.09$   & $-0.1$ & $-0.2$ & $-0.2$ \\
$[A_{1/2}]_{D1}$
& $-8.5$  & $-9.6$ & $-15.4$ & $-13.7$ \\[1em]

$[\sqrt{\Gamma_\eta} A_{3/2}]_{D1}$
& 0.7 & 0.8& 1.3& 1.1 \\
$[A_{3/2}]_{D1}$
& 64.8& 72.6& 116.5& 104.2 \\[1em]

$\alpha$
& $-1.0$   & $-1.0$ & $1.0$ & $1.0$ \\
$\beta$
& 1.1& 0.9& $-$0.3& 0.1 \\
$\delta$
& 1.7& 1.1& $-$2.7& $-$1.3 \\[1em]

$\chi^2(XS)$
& 310.5& 284.7& 287.4& 281.6 \\
$\chi^2(POL)$
& 13.2& 13.2& 14.5& 13.2\\
$\chi^2_{TOT}/DF$
& 2.2& 2.0& 2.1& 2.0 \\[1em] \hline\hline

$\xi$
& $2.19\pm 0.15$  & $2.20\pm 0.14$ & $2.20\pm 0.15$ & $2.21\pm 0.14$
\end{tabular}
\end{center}
\end{table}

\begin{table}
\caption{Fits using the  Bates data\protect\cite{dytman} only.}
\label{table11}
\begin{center}
\small
\begin{tabular}{lcccc}
$\Lambda^{2}$& 2.0 & 1.2 & 2.0 &1.2
\\[1em] \hline
$g_\eta$
& 5.9& 4.6& 5.9& 4.3 \\[1em]

$[\sqrt{\Gamma_\eta} A_{1/2}]_{S1}$
& 26.2&26.3& 26.2& 26.4 \\
$[A_{1/2}]_{S1}$ & 95.5& 96.0& 95.8& 96.2 \\[1em]

$[\sqrt{\Gamma_\eta} A_{1/2}]_{S2}$
& $-5.5$& $-6.8$& $-6.4$& $-7.4$ \\
$[A_{1/2}]_{S2}$ & 142.9& 176.7& 164.6& 192.2 \\[1em]

$[\sqrt{\Gamma_\eta} A_{1/2}]_{P1}$
& 0.02& 0.02& $0.02$& $-0.8$ \\
$[A_{1/2}]_{P1}$ & 0.1& 0.1& $0.1$& $-3.7$ \\[1em]

$[\sqrt{\Gamma_\eta} A_{1/2}]_{P2}$
& 2.1& 2.8& 3.1& 3.6 \\
$[A_{1/2}]_{P2}$ & 12.1& 16.4& 18.1& 20.7 \\[1em]

$[\sqrt{\Gamma_\eta} A_{1/2}]_{D1}$
& $-$0.5& $-0.5$& $-0.1$& $-0.1$ \\
$[A_{1/2}]_{D1}$ & $-46.8$& $-46.5$& $-9.6$ & $-9.8$ \\[1em]

$[\sqrt{\Gamma_\eta} A_{3/2}]_{D1}$
& 3.9& 3.9& 0.8& 0.8 \\
$[A_{3/2}]_{D1}$ & 355.4& 353.3& 72.9& 74.0 \\[1em]

$\alpha$
& $-1.0$   & $-1.0$ & $1.0$ & $1.0$ \\
$\beta$
& 0.7& 0.7& 8.2& 8.6 \\
$\delta$
&  $-$0.3& $-0.4$& 10.0& 10.0 \\[1em]

$\chi^2(XS)$
&  4.8& 4.7& 5.7& 5.8 \\
$\chi^2_{TOT}/DF$
& 0.6& 0.6& 0.7& 0.7  \\[1em] \hline\hline

$\xi$
& $2.19$  & $2.20$ & $2.20$ & $2.21$
\end{tabular}
\end{center}
\end{table}

\begin{table}
\caption{Contributions to the real part of the $E_{0^+}$ multipole, in
units of $10^{-3}/m_{\pi+}$, for the $\eta$ and $\pi^0$
photoproduction at their respective thresholds.
For all fits, $\alpha =+1$ and $\Lambda^{2}=1.2 GeV^{2}$ are used. Fits A, B,
C and D are defined in the text, based on different combinations of data bases.
Ref. a is the reanalysis\protect\cite{bernstein,bergstrom}
of the near threshold experiment\protect\cite{mainz}
on $\pi^\circ$ photoproduction on the proton.}
\label{table12}
\begin{center}
\begin{tabular}{lccccc}
&&&&&\\
Contributions      & Fit A &Fit B & Fit C   &Fit D&  $\pi^0$ \\[1em] \hline
&&&& \\[1em]
Nucleon Born terms &$-6.05$ &$-6.37$ &$-6.37$  & $-3.47$& $-2.46$  \\[1em]
$\rho+\omega$      &$2.89$ &$2.89$   &$2.89$ & $2.89$ & $0.04$ \\[1em]
$\Delta (1232)$    &$--$ &$--$       &$--$       &$--$ & $0.35$ \\[1em]
$N^{*}(1535)$      &$12.06$  &$12.16$  &$12.16$  &$12.16$& $0.03$ \\[1em]
$N^{*}(1440)$      &$-0.47$  & 0.08 &$-0.27$ &$-1.22$& $0.01$ \\[1em]
$N^{*}(1650)$      &$-0.94$  &$-1.00$  &$-1.00$&$-1.00$  & $0.01$ \\[1em]
$N^{*}(1520)$      &$1.46$   & 1.39&$-6.08$ &0.43& $0.08$  \\[1em]
$N^{*}(1710)$      &$ 0.25$  &0.23  &0.23&0.23    & $0.00$  \\[1em]
Total              &$9.21$     &9.39&1.56 &10.02  & $-1.94$  \\[1em]
Experiment
&$--$   &$--$  &$--$  &$--$ & $-2.0 \pm 0.2\:^{\textstyle a}$ \\[1em]
\end{tabular}
\end{center}
\end{table}


\begin{thebibliography}{999}
\bibitem{note0}
In this paper, the usual\cite{debaenst} multipole notation has been used.
\bibitem{note1}
Throughout this paper, the Bjorken and Drell conventions\cite{bjorken}
on the metric and $\gamma$ matrices are used. Also, natural units
$\hbar =c=1$ are used.
\bibitem{note2}
Since the anomalous magnetic moments couple to $F_{\mu\nu}$, gauge
invariance is always preserved in the magnetic interaction terms, even with
form factors.
\bibitem{note3}
The quoted error is due to the change of the $\chi^{2}/DF$ by 1. If
instead the total $\chi^{2}$ is used, the errors given by MINUIT are
much smaller and are not quoted here.
\bibitem{note4}
In the literature, there is  a discussion of a non-local wave equation for
spin-$3/2$ field but such an equation has been rejected because of
singularities\cite{van}.
\end{thebibliography}
\end{document}